\documentclass[12pt,letterpaper]{JHEP3}
\usepackage{axodraw}
\usepackage{amsmath}
\usepackage{amssymb}
\usepackage{cite}
\usepackage{graphicx}

\title{Multiverse Understanding of Cosmological Coincidences}

\author{Raphael Bousso, Lawrence J.~Hall and Yasunori Nomura\\
  Center for Theoretical Physics, Department of Physics\\
  University of California, Berkeley, CA 94720-7300, U.S.A.\\
  {\em and}\\
  Lawrence Berkeley National Laboratory, Berkeley, CA 94720-8162,
  U.S.A.}

\abstract{
There is a deep cosmological mystery: although dependent on very 
different underlying physics, the timescales of structure formation, 
of galaxy cooling (both radiatively and against the CMB), and of 
vacuum domination do not differ by many orders of magnitude, but 
are all comparable to the present age of the universe.  By scanning 
four landscape parameters simultaneously, we show that this quadruple 
coincidence is resolved.  We assume only that the statistical 
distribution of parameter values in the multiverse grows towards 
certain catastrophic boundaries we identify, across which there 
are drastic regime changes.  We find order-of-magnitude predictions 
for the cosmological constant, the primordial density contrast, the 
temperature at matter-radiation equality,  the typical galaxy mass, 
and the age of the universe, in terms of the fine structure constant 
and the electron, proton and Planck masses.  Our approach permits 
a systematic evaluation of measure proposals; with the causal patch 
measure, we find no runaway of the primordial density contrast and 
the cosmological constant to large values.
}

\begin{document}

\section{Introduction}
\label{sec:intro}

In the evolution of the universe there are certain critical time scales 
that separate one era from another.  The electroweak phase transition 
occurred at a time of order $10^{-12}~{\rm sec.}$, the QCD phase transition 
at order $10^{-4}~{\rm sec.}$ and the transition from radiation to matter 
domination at order $10^{12}~{\rm sec.}$  The time scale of inflation 
is very uncertain, and could be as long as a second or as short as 
$10^{-38}~{\rm sec.}$  Apparently the universe evolved smoothly for 
many decades in any given era, but between these eras, at well separated 
time scales, there were dramatic changes of regime.

This picture breaks down in a remarkable way during the present era. 
There are several time scales that originate from very different 
fundamental physics, and would be expected to differ by many orders 
of magnitude, but all have the same order of magnitude.  Each of these 
time scales by themselves could herald a change of regime.  The change 
from matter to vacuum domination, instigating a new era of accelerated 
expansion, occurs at $t_\Lambda \approx \Lambda^{-1/2}$, where 
$\Lambda = 8\pi G_{\rm N} \rho_\Lambda$ is the cosmological constant. 
Another change of regime occurs at the time $t_{\rm vir}$ which is 
determined by a combination of the size of the primordial density 
perturbations and the temperature of matter radiation equality.  At 
this time, density perturbations become non-linear.  The universe 
transits out of the nearly homogeneous state it had for a number of 
eras and becomes clumpy.  There are two possibilities for subsequent 
evolution.  One is that the clumps or proto-galaxies cool very 
quickly to form galaxies, $t_{\rm cool} \ll t_{\rm vir}$, followed 
by fragmentation and star formation.  In this case the change at 
$t_{\rm vir}$ is from a regime of approximate homogeneity to one 
of star formation.  The second possibility is that $t_{\rm cool} 
\gg t_{\rm vir}$, so that there is first an era of virialized 
proto-galaxies, and then a much later era during which they cool. 
In fact, two very different processes can contribute to cooling: 
radiative emission and inverse Compton scattering from the background 
radiation, with time scales $t_{\rm rad}$ and $t_{\rm comp}$, 
respectively, and the cooling time scale $t_{\rm cool}$ is given 
by the smaller of these two.  Finally, once proto-galaxies have 
cooled, they give rise to additional time scales, including the 
time scale for the subsequent appearance of observers, $t_{\rm obs}$, 
and the stellar burning time scale $t_{\rm burn}$.

Other critical time scales in the past history of our universe span 
more than $60$ orders of magnitude.  Those of future phenomena, such 
as the decay of matter, black hole evaporation, etc., can range over 
thousands of orders of magnitude~\cite{Dyson:1979zz}.  One might expect 
that the five time scales $t_\Lambda$, $t_{\rm vir}$, $t_{\rm rad}$, 
$t_{\rm comp}$ and $t_{\rm obs}$, dependent as they are on entirely 
different physics, would also be well separated.  Remarkably, they 
are all comparable,
\begin{equation}
  t_\Lambda \sim t_{\rm vir} \sim t_{\rm rad} 
    \sim t_{\rm comp} \sim t_{\rm obs},
\label{eq:coscoinc}
\end{equation}
having an order of magnitude of $10^{10}~{\rm years}$.%
\footnote{As a by-product of understanding these coincidences, we 
 will find that the time scale for a main sequence star of maximal 
 mass, $t_{\rm burn}$, is constrained to be parametrically equal 
 to these other time scales.}
The picture of well separated eras breaks down during the present time, 
since many changes in regime are occurring simultaneously.  It is hard 
to accept that Eq.~(\ref{eq:coscoinc}) is purely coincidental; but it 
is also hard to see how it could be derived from symmetries of some 
underlying field theory.

If $\Lambda$ varies in the multiverse, one of the coincidences of 
Eq.~(\ref{eq:coscoinc}) can be explained by environmental selection. 
Only those universes with $t_\Lambda > t_{\rm vir}$ have any large 
scale structure and can be observed~\cite{Weinberg:1987dv} (see 
also \cite{Barrow-Tipler}).  Given that most universes have a large 
cosmological constant, the statistical prediction over the multiverse 
is $t_\Lambda \sim t_{\rm vir}$.  This result seems particularly 
important since it is the only known solution to the cosmological 
constant problem, and it predicted the discovery of nonzero vacuum 
energy.  In fact the observed value of the cosmological constant is 
somewhat smaller than expected from this result~\cite{Efstathiou:1995ne}, 
but it is important to note that the prediction depends on the measure 
used to define multiverse probability distributions.  Using the 
causal patch measure~\cite{Bousso:2006ev}, the prediction is 
modified to $t_\Lambda \sim t_{\rm vir} + t_{\rm cool} + t_{\rm obs}$, 
and is more successful~\cite{Bousso:2007kq}.

It is important to understand that it is legitimate in such 
calculations to hold all parameters but one fixed, even though many 
parameters can vary in the string landscape.  After all, it yielded 
a first, nontrivial prediction that could have failed (e.g., if 
$\Lambda$ had not been discovered).  Once additional parameters are 
allowed to vary simultaneously, the theory has another opportunity to 
fail, or to succeed.  A long-standing concern about the prediction for 
the cosmological constant is that simultaneous scanning of the virial 
density leads to runaway behavior favoring large values of both 
quantities.  However, runaway behavior need not ensue if the range 
of new scanning parameters is constrained by additional catastrophic 
boundaries, and if multiverse forces point towards these boundaries.

In this paper, we study this possibility.  We identify a number 
of possible catastrophic boundaries.  We argue that under suitable 
assumptions about the prior distribution of parameters in the 
landscape, these boundaries can prevent runaway.  What makes this 
viewpoint especially compelling is that it explains all of the 
coincidences in Eq.~(\ref{eq:coscoinc}), allowing for novel 
cosmological predictions from environmental selection.

\paragraph{Outline}
In section~\ref{sec:multi_force}, we review how predictions arise 
from multiverse probability distributions that grow towards catastrophic 
boundaries, across which there are significant changes in the anthropic 
weighting factor~\cite{Hall:2007ja}.  We stress the utility of the 
multiverse force, the logarithmic derivative of the distribution 
function, and of a set of variables chosen to be orthogonal to the 
catastrophic boundaries.  In section~\ref{sec:astro} we review analytic 
estimates for a number of astrophysical processes: halo virialization, 
halo disruption by a cosmological constant, radiative cooling, and 
inverse Compton cooling of proto-galaxies.  Our aim is to obtain 
an approximate parametric dependence of astrophysical scales in our 
universe on parameters such as the fine structure constant, $\alpha$, 
the electron mass, $m_e$, the proton mass, $m_p$, and $\Lambda$, 
which may vary in the multiverse.

In sections~\ref{sec:pred_cool}, \ref{sec:CC-1}, \ref{sec:CC-2} and 
\ref{sec:comp} we successively allow the virial density, the cosmological 
constant, the time scale of observers, and the Compton cooling time 
scale to scan.  In each section, we add one more scanning parameter 
and one more catastrophic boundary, and we obtain one new prediction, 
while maintaining the previous predictions and results.  At each stage, 
we derive the range in which multiverse forces would have to lie so 
that our universe is not atypical.

In section~\ref{sec:pred_cool}, we assume that the matter density at 
virialization scans in the multiverse.  We identify a catastrophic 
boundary: for sufficiently small density, proto-galaxies cannot cool. 
Assuming a multiverse force towards small density, we find predictions 
for the virial density and the mass of a galaxy,
\begin{eqnarray}
  && \rho_{\rm vir} \sim 
  \frac{G_{\rm N} m_e^4 m_p^2}{\alpha^4},
\label{eq:rho_vir-pred_intro}\\
  && M_{\rm gal} \sim \frac{\alpha^5}{G_{\rm N}^2 m_e^{1/2} m_p^{5/2}},
\label{eq:M_gal-pred_intro}
\end{eqnarray}
in terms of $\alpha$, $m_e$, $m_p$, and the Newton constant, $G_{\rm N}$. 
(We have omitted numerical coefficients that are displayed in the main 
text.)  This also explains the coincidence that the timescales for 
radiative cooling and galactic halo virialization are comparable:
\begin{equation}
  t_{\rm rad} \sim t_{\rm vir}.
\label{eq:rad=vir}
\end{equation}
The distance of the observed virial density from the catastrophic 
boundary is used to place empirical bounds on the strength of the 
multiverse force.

Next, in section~\ref{sec:CC-1}, we consider scanning the cosmological 
constant $\Lambda$ in addition to the virial density.  In a certain 
class of measures that includes the scale factor cutoff measure proposed 
in Ref.~\cite{DeSimone:2008bq}, $\Lambda$ is constrained from above by 
the disruption of halo formation.  There is a potential runaway favoring 
large values of both quantities, since a larger virial density allows 
for a larger cosmological constant while preserving structure formation. 
However, for a certain range of multiverse force directions, this 
runaway is avoided and one predicts that typical values of the two 
parameters are close to the intersection of the two boundaries. 
This leads to a second prediction, now for the vacuum energy:
\begin{equation}
  \rho_\Lambda  \sim \frac{1}{G_{\rm N} t_\Lambda^2} \sim 
    \frac{G_{\rm N} m_e^4 m_p^2}{\alpha^4}.
\label{eq:rho_Lambda-pred_intro}
\end{equation}
It explains, moreover, the double coincidence
\begin{equation}
  t_\Lambda \sim t_{\rm rad} \sim t_{\rm vir}.
\label{eq:Lambda=rad=vir}
\end{equation}
As in the previous section, we derive empirical constraints on the 
values of multiverse force components by comparing the observed 
$\rho_\Lambda$ and $\rho_{\rm vir}$ to the critical values.  We can 
compare these constraints to theoretical predictions, which depend 
on the measure on the multiverse, and on the prior distribution in 
the landscape (which is known for $\rho_\Lambda$).  For the measure 
of~\cite{DeSimone:2008bq}, the empirical and theoretical limits 
conflict.  This illustrates how our approach can be used as 
a systematic discriminator between measures.

In section~\ref{sec:CC-2}, we consider a different class of 
measures, where the weight of observations made after $t_\Lambda$ 
is exponentially suppressed.  (This class includes the 
causal patch measure~\cite{Bousso:2006ev}, the causal diamond 
measure~\cite{Bousso:2007kq}, and the modified scale factor measure 
of~\cite{Bousso:2008hz}.)  In the sharp boundary approximation, 
catastrophe occurs unless $t_\Lambda > t_{\rm vir} + t_{\rm obs}$. 
This condition is approximately equivalent to {\em two\/} 
boundaries---the upper bound on the cosmological constant 
that appeared in the previous section, and a new boundary 
$t_{\rm obs} < t_\Lambda$.  Correspondingly, this class of measures 
allows us to scan a third parameter, $t_{\rm obs}$, in addition 
to scanning $\rho_{\rm vir}$ and $\rho_\Lambda$.  A wide range 
of multiverse forces will favor universes close to the intersection 
of the three catastrophic boundaries, allowing an explanation 
of the triple coincidence
\begin{equation}
  t_{\rm obs} \sim t_\Lambda \sim t_{\rm rad} \sim t_{\rm vir}.
\label{eq:obs=Lambda=rad=vir}
\end{equation}
We determine quantitative empirical constraints on force components 
and find that they are perfectly compatible with the prior distribution 
of $\rho_\Lambda$ if the causal patch measure is used.

Halos that virialize in a sufficiently hot universe cool by inverse 
Compton scattering against the cosmic background radiation.  We 
do not know whether this change of regime is catastrophic, but in 
section~\ref{sec:comp}, we explore the consequences of assuming 
that it is.  With this new boundary, we can allow a fourth parameter 
to scan, which we choose to be the strength of primordial density 
perturbations $Q$.  In this case a further prediction follows from 
assuming a multiverse distribution that prefers large values of $Q$
\begin{equation}
  Q \sim \frac{m_e}{m_p}.
\label{eq:Q0-pred_intro}
\end{equation}
Since $Q$ and $\rho_{\rm vir}$ determine the temperature at equality, 
combining this with Eq.~(\ref{eq:rho_vir-pred_intro}) yields
\begin{equation}
  T_{\rm eq} \sim \frac{1}{\alpha} (G_{\rm N} m_e m_p)^{1/4} m_p.
\label{eq:Teq-pred_intro}
\end{equation}
This additional prediction, combined with Eq.~(\ref{eq:obs=Lambda=rad=vir}), 
explains the quadruple coincidence
\begin{equation}
  t_{\rm comp} \sim t_{\rm obs} \sim t_\Lambda 
    \sim t_{\rm rad} \sim t_{\rm vir}.
\label{eq:comp=obs=Lambda=rad=vir}
\end{equation}

In section~\ref{sec:concl}, we discuss how far this program of adding 
more scanning parameters and catastrophic boundaries can be taken.  We 
argue that many more predictions are possible, and we consider various 
ultimate solutions to runaway behavior.

\section{A Multiverse Force and Unnatural Predictions}
\label{sec:multi_force}

In this section, we review the approach to multiverse predictions 
proposed in Ref.~\cite{Hall:2007ja}.  We will also discuss the 
quantitative relation between the strength of a multiverse force 
and the proximity of typical observers to the corresponding 
catastrophic boundary.

A landscape of vacua, such as the string landscape, will contain 
scanning parameters, $x$, that vary between vacua~\cite{Bousso:2000xa}. 
We can consider any parameters that enter physical phenomena; they 
need not be Lagrangian parameters.  The multiverse distribution
\begin{equation}
  dN = f(x)\, d\ln x
\label{eq:dN}
\end{equation}
describes the relative number of times the value $x$ is observed, 
and thus its probability.  In most situations, the distribution
\begin{equation}
  f(x) = \tilde{f}(x) w(x)
\label{eq:f-x}
\end{equation}
factorizes into an a priori distribution, $\tilde{f}(x)$, of vacua 
in the theory landscape and an anthropic weighting factor, $w(x)$, 
roughly the number of observers in vacua where the parameter takes 
on the value $x$.  (This number naively vanishes or diverges for all 
$x$ because of eternal inflation, and a cutoff or measure is required 
to make it well-defined.)

\paragraph{Living on the edge}
In general, to make a prediction one needs to understand both 
$\tilde{f}(x)$ and $w(x)$---the landscape, the measure, and the 
abundance of observers---in great detail.  Under the following 
two conditions, however, quantitative predictions can be made even 
though only qualitative features of $\tilde{f}$ and $w$ are known:
\begin{itemize}
\item There are catastrophic boundaries where $w(x)$ drops sharply. 
  Studying physics in the parameter space of $x$, one often discovers 
  that there are boundaries in this space across which there is a 
  sudden change in the physics.  For example, this could be a phase 
  transition or the stability criterion for some complex object.  In 
  the extreme limit, $w(x)$ can be represented by a theta function, so 
  that observers are found in universes on one side of the boundary, 
  but not in universes on the other side.  This limit need not be 
  exactly realized in practice, but it will simplify our analysis 
  considerably.
\item In the region of $x$ where observers are possible, the 
  probability distribution, $f(x)$, increases towards the catastrophic 
  boundary(s).  Thus, most universes do not support observers.
\end{itemize}
In this situation, a typical observer would expect to measure $x$ 
close to the boundary.  Vacua with very different values of $x$ either 
contain no observers, or are very rare in the multiverse.

\begin{figure}[t]
\begin{center}
\begin{picture}(250,47)(-5,-22)
  \Line(-10,0)(250,0) \Line(250,0)(243,4)  \Line(250,0)(243,-4)
  \Text(250,-8)[t]{\large $x$}
  \Line(180,-10)(180,10) \Text(184,-16)[t]{\large $x_b$}
  \Line(180.0,5.0)(185.0,10.0)
  \Line(180.0,0.0)(190.0,10.0)
  \Line(180.0,-5.0)(190.0,5.0)
  \Line(180.0,-10.0)(190.0,0.0)
  \Line(185.0,-10.0)(190.0,-5.0)
  \Line(65.0,13.1)(104.8,13.1) \Line(65.0,12.9)(104.8,12.9)
  \Line(105,13)(98,16) \Line(105,13)(98,10)
  \Text(60,13)[r]{\large ${\bf F}$}
  \Vertex(167,0){3} \Text(169,-8)[t]{\large $x_o$}
\end{picture}
\caption{A strong multiverse force towards a catastrophic boundary at 
 $x_b$ implies that a typical observer will be located close to the 
 boundary at $x_o$.}
\label{fig:1-dim}
\end{center}
\end{figure}
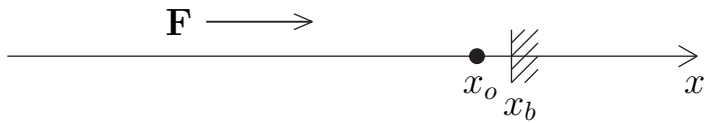

The crucial point is that this quantitative prediction depends mostly 
on the location of the boundary, which can be calculated using the 
conventional law of physics.  It depends only weakly on the multiverse, 
in that only the above qualitative assumptions on the multiverse 
distribution must be satisfied.  This situation is illustrated in 
Fig.~\ref{fig:1-dim} for the case of a single scanning parameter 
with a catastrophic boundary at $x = x_b$.

The prediction $x \sim x_b$ does not follow from any symmetry of 
the theory; indeed it may appear coincidental.  For example, it may 
relate two time scales or two mass scales that do not have a common 
origin.  Since there is no symmetry reason that the observed value 
$x_o$ should be close to $x_b$, from the conventional viewpoint the 
multiverse predictions appear quite unnatural.

Taking $x$ to be the vacuum energy density, $\rho_\Lambda = 
\Lambda/8\pi G_{\rm N}$, Fig.~\ref{fig:1-dim} illustrates the prediction 
of the cosmological constant, with $x_b \approx \rho_{\rm vir}$. 
Using a logarithmic scale, the multiverse distribution indeed 
grows towards the boundary.  This represents the cosmological 
constant problem.  The naturalness probability, defined by $P = 
|\int_{x_o}^{x_b}\! f(x)\, dx/\int\!  f(x)\, dx|$, is of order 
$10^{-120}$, and from the usual symmetry viewpoint, it is extremely 
unnatural for $\rho_\Lambda$ to take the observed value.  In the 
multiverse picture, however, the vast majority of universes are 
simply irrelevant since they contain no large scale structure and 
thus no observers.  Observing $\rho_\Lambda \sim \rho_{\rm vir}$, 
therefore, is not a mystery.

\paragraph{Force strength}
The steepness of the distribution function $f(x)$ determines how 
close a typical observer lies to the boundary.  We can refine our 
analysis by defining the multiverse probability force
\begin{equation}
  F = \frac{\partial \ln f}{\partial \ln x}.
\label{eq:force}
\end{equation}
In this paper we will consider only power law distributions, $f(x) 
\propto x^p$.  Then the force is simply given by the power, $F = p$. 
Defining $x_\nu$ to be the point such that the fraction of observers 
farther from the boundary than $x_\nu$ is $\nu$, we find
\begin{equation}
  \frac{x_\nu}{x_b} = \nu^{1/p}.
\label{eq:x_n}
\end{equation}
Thus $x_{0.5}/x_b$ is a useful measure of the typical ``distance'' 
of an observer from the boundary; it is $1/2$ for $p=1$ and rapidly 
becomes smaller (larger) as $p$ decreases (increases), since $1/p$ 
appears in the exponent in Eq.~(\ref{eq:x_n}).  For $p=0$ the 
distribution is logarithmic, with equal population of all logarithmic 
intervals.  In this case, one does not expect to find $x$ close to 
$x_b$, so it is of no interest here.

If the force is somehow known, then we can predict not only that $x$ 
should be close to $x_b$, but how close it should be.  Roughly, one 
expects to measure a value of order $x_{0.5}$, and one would be mildly 
surprised to measure values farther from the boundary than, say, 
$x_{0.16}$, or closer than $x_{0.84}$.

In practice, the strength of the force is not precisely known from the 
theory of the landscape.  We can, however, still argue that the multiverse 
is likely playing a role in determining the value of a parameter $x$ if 
$x_o$ is anomalously close to $x_b$, since such a phenomenon is extremely 
hard to explain in any other framework.  In this circumstance, one can 
reverse the logic given above and derive bounds on the strength of the 
force; for example, if $x_o$ is extremely close to $x_b$, the force is 
likely to be quite strong.  This information can then be used to study if 
a certain assumption on the landscape or measure is consistent with data.

Let us make this point more precise.  Given $x_o$ and $x_b$, the observed 
and boundary values of $x$, Eq.~(\ref{eq:x_n}) allows us to compute 
the force $p$ such that a fraction $\nu$ of observers are farther 
from the boundary than we are:
\begin{equation}
  p(\nu) = \frac{\ln \nu}{\ln (x_o/x_b)}.
\label{eq:force-p}
\end{equation}
For example, suppose that we are willing to reject models that would 
put us farther than one standard deviation from the median.  Then we 
can infer that the multiverse force lies in the range $p(0.16)$ to 
$p(0.84)$.  If we allow two standard deviations, the force must lie 
in the range $p(0.025)$ to $p(0.975)$.

As we shall see below, such empirical bounds on the multiverse force 
can be quite useful, for two reasons.  First, there may be more than 
one way of obtaining a bound on the multiverse force acting on a 
particular parameter, allowing for a nontrivial consistency check. 
Second, even though we may not be able to compute the multiverse 
force from first principles, there may be theoretical arguments why 
it should satisfy a certain upper or lower bound.  In these cases, 
comparison of the theoretical arguments to the empirical bounds offers 
a possibility of falsifying various assumptions on the properties 
of the landscape and the measure.

\begin{figure}[t]
\begin{center}
\begin{picture}(250,172)(-5,-13)
  \Line(-10,0)(250,0) \Line(250,0)(243,4)  \Line(250,0)(243,-4)
  \Line(0,-10)(0,150) \Line(0,150)(-4,143) \Line(0,150)(4,143)
  \Text(250,-8)[t]{\large $x_1$} \Text(-8,147)[r]{\large $x_2$}
  \CArc(-60,-252)(395,44,77)   \Text(26,143)[]{\large $A$}
  \CArc(-126,937)(930,280,293) \Text(30,11)[]{\large $B$}
  \Line(32.0,132.0)(42.0,142.0)
  \Line(40.0,130.0)(50.0,140.0)
  \Line(47.9,127.9)(57.9,137.9)
  \Line(55.6,125.6)(65.6,135.6)
  \Line(63.2,123.2)(73.2,133.2)
  \Line(70.7,120.7)(80.7,130.7)
  \Line(78.0,118.0)(88.0,128.0)
  \Line(85.2,115.2)(95.2,125.2)
  \Line(92.3,112.3)(102.3,122.3)
  \Line(99.3,109.3)(109.3,119.3)
  \Line(106.2,106.2)(116.2,116.2)
  \Line(113.0,103.0)(123.0,113.0)
  \Line(119.7,99.7)(129.7,109.7)
  \Line(126.2,96.2)(136.2,106.2)
  \Line(132.7,92.7)(142.7,102.7)
  \Line(139.0,89.0)(149.0,99.0)
  \Line(145.3,85.3)(155.3,95.3)
  \Line(151.5,81.5)(161.5,91.5)
  \Line(157.5,77.5)(167.5,87.5)
  \Line(163.5,73.5)(173.5,83.5)
  \Line(169.4,69.4)(179.4,79.4)
  \Line(175.2,65.2)(185.2,75.2)
  \Line(180.9,60.9)(190.9,70.9)
  \Line(183.7,53.7)(196.5,66.5)
  \Line(192.0,52.0)(209.7,69.7)
  \Line(197.4,47.4)(207.4,57.4)
  \Line(211.3,61.3)(226.3,76.3)
  \Line(202.7,42.7)(212.7,52.7)
  \Line(208.0,38.0)(218.0,48.0)
  \Line(213.1,33.1)(223.1,43.1)
  \Line(218.2,28.2)(228.2,38.2)
  \Line(164.7,44.7)(179.7,58.7)
  \Line(148.1,38.1)(163.1,53.1)
  \Line(133.5,33.5)(148.5,48.5)
  \Line(119.2,29.2)(134.2,44.2)
  \Line(105.2,25.2)(120.2,40.2)
  \Line(91.6,21.6)(106.6,36.6)
  \Line(78.3,18.3)(93.3,33.3)
  \Line(65.2,15.2)(80.2,30.2)
  \Line(52.4,12.4)(67.4,27.4)
  \Line(39.8,9.8)(54.8,24.8)
  \Line(35,94.1)(73,106.1) \Line(35,93.9)(73,105.9)
  \Line(73,106)(65.4,106.8) \Line(73,106)(67.2,101.0)
  \Text(30,92)[r]{\large ${\bf F}$}
  \DashCArc(-60,-263)(395,60,68){2}
  \Line(137.5,79.1)(133.0,85.3) \Line(137.5,79.1)(130.0,80.1)
  \Vertex(166,62){3}
\end{picture}
\caption{With two scanning parameters, $(x_1,x_2)$, a multiverse force 
 towards a single catastrophic boundary, such as $A$, may lead to runaway 
 behavior.  The introduction of a second boundary, $B$, can prevent this 
 runaway and lead to predictions for both parameters.}
\label{fig:2-dim}
\end{center}
\end{figure}
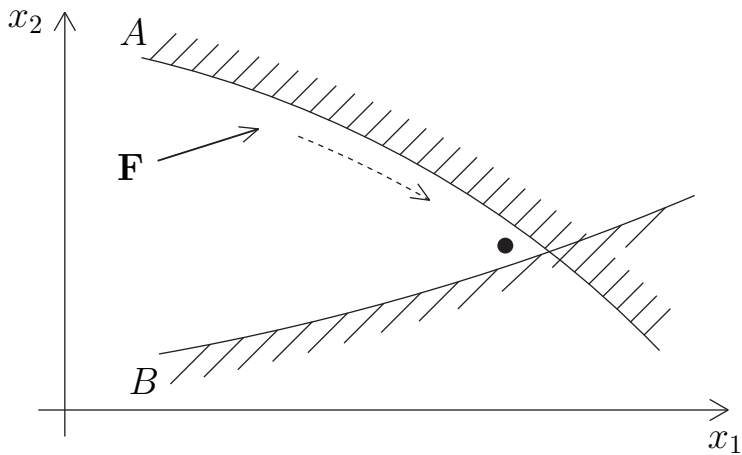

\paragraph{Multiple parameters}
Suppose that two parameters scan.  Consider, for example, the 
catastrophic boundary labeled $A$ in Fig.~\ref{fig:2-dim}.
In general the multiverse force is not perpendicular to the boundary, 
so there is now a runaway problem, as shown by the dashed arrow.  This 
runaway behavior may be halted in several ways.  A special shape of the 
boundary could prevent runaway, but for many boundaries this is not the 
case.  The runaway could be prevented by $f(x)$ reaching a maximum at 
some point along the boundary; however, in this case this special point 
on the boundary results from an assumption on the form of $f(x)$ and 
is not predicted.

Alternatively, the runaway problem can be solved by introducing a 
second catastrophic boundary, as illustrated by the boundary $B$ in 
Fig.~\ref{fig:2-dim}.  In this case, the boundaries define a special 
point by their intersection, and a multiverse force in a wide range of 
directions will lead to typical observers being close to this point, 
as shown by the dot in Fig.~\ref{fig:2-dim}.  Not only does the second 
boundary stop the runaway behavior, it leads to a second prediction, 
which will again appear coincidental and unnatural from a symmetry 
viewpoint.  Typical observers are in universes near the ``tip of the 
cone'', a special point in {\em two-dimensional\/} parameter space.

With $n$ scanning parameters there could be many runaway directions. 
But with $n$ boundaries of codimension one, it is possible that all 
the scanning parameters are determined.  With the multiverse force 
pointing in a generic subset of directions, the most probable observed 
universes will be at the tip of an $n$-dimensional cone, giving $n$ 
predictions.

In general, these predictions will relate the scanning parameters 
$x$ to fixed parameters $y$, which are not allowed to scan---either 
because they do not vary in the landscape, or because we choose to 
consider only a subset of the landscape.  It is possible that, for 
the phenomena of interest, all parameters scan, and the configuration 
of boundaries together with our assumptions on $f(x)$ prevent any 
runaway and determine all the underlying parameters.  In practice 
it may be hard to come up with such a closed, completely determined 
set of parameters, and the multiverse may not lead to such situations. 
Still, it is worth exploring how far this program of multiverse 
predictions can be taken.

In this paper, we consider mainly power law forces, $f(x_1,\ldots,x_n) 
\propto \prod_{i=1}^n x_i^{p_i}$, and power law catastrophic boundaries, 
$\prod_{j=1}^n x_j^{b_{ij}} = \beta_i$, $i=1,\ldots,n$, which generically 
intersect at a critical point $(x_{1,{\rm c}},\ldots,x_{n,{\rm c}})$. 
It will be convenient to work with the logarithmic quantities
\begin{equation}
  z_i \equiv \ln\frac{x_i}{x_{i,{\rm c}}},
\label{eq:zx}
\end{equation}
and to use capital letters $Z$, $P$, and $B$, to denote the vectors 
and matrix constructed from $z_i$, $p_i$, and $b_{ij}$.  Then the 
critical point is at $Z=0$.

\begin{figure}[t]
\begin{center}
\begin{picture}(395,157)(-100,-18)
  \Line(-25,0)(205,0) \Line(205,0)(198,4) \Line(205,0)(198,-4)
  \Line(-15,-10)(-15,130) \Line(-15,130)(-19,123) \Line(-15,130)(-11,123)
  \Text(-22,128)[r]{\large $\ln x_2$}
  \Text(203,-8)[t]{\large $\ln x_1$}
  \Line(-10,15)(105,130)
  \Line(-10,15)(-7.90,22.72) \Line(-10,15)(-2.28,17.10)
  \Text(-5,11)[t]{\large $u_1$}
  \Line(-3,22)(-3,36)
  \Line(2,27)(2,41)
  \Line(7,32)(7,46)
  \Line(12,37)(12,51)
  \Line(17,42)(17,56)
  \Line(22,47)(22,61)
  \Line(27,52)(27,66)
  \Line(32,57)(32,71)
  \Line(37,62)(37,76)
  \Line(42,67)(42,81)
  \Line(47,72)(47,86)
  \Line(52,77)(52,91)
  \Line(57,82)(57,96)
  \Line(77,102)(77,116)
  \Line(82,107)(82,121)
  \Line(87,112)(87,126)
  \Line(92,117)(92,130)
  \Line(97,122)(97,130)
  \Line(102,127)(102,130)
  \Line(62,87.0)(62,109.5)
  \Line(67,92.0)(67,107.0)
  \Line(72,91.5)(72,111.0)
  \Line(-5,130)(205,25)
  \Line(205,25)(197.01,24.56) \Line(205,25)(200.56,31.66)
  \Text(206,20)[t]{\large $u_2$}
  \Line(-3,129.0)(-3,130)
  \Line(2,126.5)(2,130)
  \Line(7,124.0)(7,130)
  \Line(12,121.5)(12,130)
  \Line(17,119.0)(17,130)
  \Line(22,116.5)(22,129.5)
  \Line(27,114.0)(27,127.0)
  \Line(32,111.5)(32,124.5)
  \Line(37,109.0)(37,122.0)
  \Line(42,106.5)(42,119.5)
  \Line(47,104.0)(47,117.0)
  \Line(52,101.5)(52,114.5)
  \Line(57,99.0)(57,112.0)
  \Line(77,89.0)(77,102.0)
  \Line(82,86.5)(82,99.5)
  \Line(87,84.0)(87,97.0)
  \Line(92,81.5)(92,94.5)
  \Line(97,79.0)(97,92.0)
  \Line(102,76.5)(102,89.5)
  \Line(107,74.0)(107,87.0)
  \Line(112,71.5)(112,84.5)
  \Line(117,69.0)(117,82.0)
  \Line(122,66.5)(122,79.5)
  \Line(127,64.0)(127,77.0)
  \Line(132,61.5)(132,74.5)
  \Line(137,59.0)(137,72.0)
  \Line(142,56.5)(142,69.5)
  \Line(147,54.0)(147,67.0)
  \Line(152,51.5)(152,64.5)
  \Line(157,49.0)(157,62.0)
  \Line(162,46.5)(162,59.5)
  \Line(167,44.0)(167,57.0)
  \Line(172,41.5)(172,54.5)
  \Line(177,39.0)(177,52.0)
  \Line(182,36.5)(182,49.5)
  \Line(187,34.0)(187,47.0)
  \Line(192,31.5)(192,44.5)
  \Line(197,29.0)(197,42.0)
%
  \DashLine(30,5)(101.67,76.67){3}
  \DashLine(80,5)(135,60){3}
  \DashLine(130,5)(168.33,43.33){3}
%
  \DashLine(41.67,66.67)(165,5){3}
  \DashLine(15,40)(85,5){3}
\end{picture}
\caption{Given $n$ boundary hyperplanes, it is convenient to work with 
 new variables $u_i$ such that the allowed region is a quadrant and the 
 critical point is at the origin of ${\mathbf R}^n$.  In these variables, 
 the distance from each boundary $u_i$ is inversely related to the 
 corresponding multiverse force component $s_i$, with no mixing between 
 components.}
\label{fig:multi}
\end{center}
\end{figure}
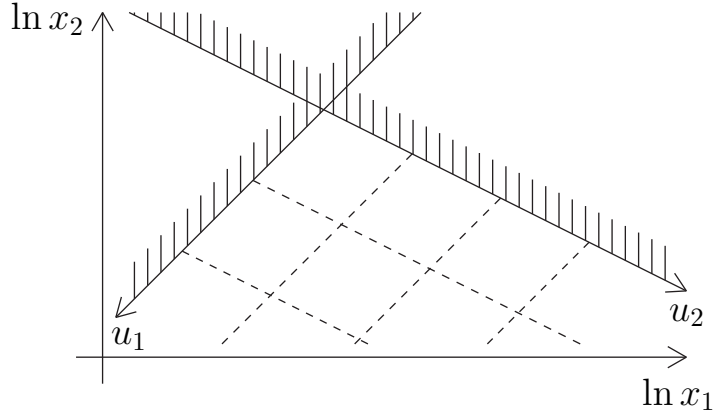

Let us define new variables $u_i \equiv b_{ij} z_j$, or
\begin{equation}
  U \equiv B Z,
\label{eq:UBZ}
\end{equation}
which measure the distances orthogonal to the $n$ boundaries.  The 
boundaries are defined by $u_i=0$, $i=1,\ldots, n$.  This is illustrated 
in the two-dimensional case in Fig.~\ref{fig:multi}.
Without loss of generality, we will choose signs such that $u_i>0$ 
in the allowed region.  The variables $U$ are convenient to work with 
because the power-law form of the distribution function is preserved 
if one or more of the $u_i$ are integrated out.  This is not generally 
the case for the $x_i$ or $z_i$.

With the new notation and variables, the distribution function takes 
the form
\begin{equation}
  \ln f(Z) = P \cdot Z = P \cdot B^{-1} U = S \cdot U,
\label{eq:fz}
\end{equation}
where $B^{-1}$ denotes the inverse matrix, and we have defined
\begin{equation}
  S \equiv (B^{-1})^T P.
\label{eq:SBP}
\end{equation}
The force vector, with components $p_i = 
\frac{\partial\ln f}{\partial z_i}$ in the original coordinates, 
has components $s_i = \frac{\partial\ln f}{\partial u_i}$ orthogonal 
to the boundaries.  Importantly, $s_i$ is the effective one-dimensional 
force on the variable $u_i$ if all other $u_j$, $j\neq i$, are 
integrated out.  (Note that the analogous statement would not 
hold true for the original forces and variables $p_i$ and $u_i$.)

To see this, consider the one-dimensional distribution functions 
$\hat{f}_i(u_i) \equiv \frac{dN}{du_i}$, which can be derived from the 
multivariate distribution by integrating out all but one variable:
\begin{equation}
  \hat{f}_i(u_i) \propto \int_0^\infty 
    \prod_{j\neq i} du_j\, |\det B|^{-1} \exp(S\cdot U) 
  \propto \exp(s_i u_i).
\label{eq:fu-integ}
\end{equation}
The one-dimensional effective force on $u_i$ is
\begin{equation}
  \frac{d\ln\hat{f}_i}{du_i}=s_i,
\label{eq:eff-s-1dim}
\end{equation}
as claimed.  With our convention that the $u_i$ are positive in the 
allowed region, the $s_i$ must all be negative to avoid runaway.

The fraction of observers farther than $u_i$ from the boundary 
$u_i = 0$ is
\begin{equation}
  \nu_i = \exp(s_i u_i).
\label{eq:far-ui}
\end{equation}
If the multiverse force is known, the characteristic distances $u_i$ 
can be predicted in analogy to Eq.~(\ref{eq:x_n}).  For example, from 
$\nu_i = 0.5$ one obtains the median values $u_i = s_i^{-1} \ln 0.5$. 
With $Z = B^{-1} U$ and Eq.~(\ref{eq:zx}), this translates into predictions 
for characteristic values of the original parameters $x_i$.

Conversely, if the multiverse force is not known, it can be constrained 
by demanding that the observed values not be improbably close to or 
far from the boundary, as in Eq.~(\ref{eq:force-p}):
\begin{equation}
  s_i(\nu) = \frac{\ln \nu}{u_{i,o}}.
\label{eq:snu}
\end{equation}
These $n$ constraints can be translated into constraints on the $n$ 
original force components $p_i$, with $P = B^T S$.

The region of parameter space that allows observers -- the observer 
region -- is assumed to be open in the above analysis.  However, 
there could be more catastrophic boundaries that make the observer 
region finite.  If these additional boundaries are not distant 
from the point describing our universe, the analysis would change 
accordingly.

\section{Parametric Dependence of Astrophysical Scales}
\label{sec:astro}

In this section, we will discuss astrophysics in our universe.  We 
will ask how astrophysical quantities, like the virial density and 
temperature, and the range of galactic masses, are related to 
``fundamental'' parameters such $\alpha$, $m_e$, $m_p$, $Q$, etc. 
Such calculations have nothing to do with the landscape or with 
imposing anthropic cuts; they are merely derivations of observed 
macroscopic phenomena from observed microphysics.  A large enough 
computer could perform them with arbitrary accuracy, given the 
observed values of the input parameters.

We will not attempt a very detailed analysis.  Since we will be mainly 
interested in the parametric dependence on the fundamental parameters, 
our goal is to capture the dominant physics that controls the phenomena 
we study.  In order to compensate for analytical shortcuts and bring 
our simple models into precise agreement with data, we will insert 
suitable order one factors where necessary (for example, $c_{\rm ion}$ 
and $c_{\rm vir}$ below).

Note that this does not amount to putting in the answer to the central 
problem investigated in this paper.  Our goal is to {\em explain\/} why 
certain combinations of fundamental parameters (namely those that appear 
in derived astrophysical and cosmological quantities) take on the values 
they do.  This question becomes meaningful only in the larger setting 
of the multiverse, where what would otherwise be fixed input parameters 
like $\alpha$, $m_e$, $m_p$, $Q$, etc., can vary.  It will be 
considered in the following sections.

\subsection{Halo formation}
\label{subsec:halo}

Complex structures, such as stars, will not form unless the virialized 
gas inside dark matter halos is able to cool.  In our universe, 
there is both a maximum and a minimum characteristic halo mass for 
which radiative cooling is successful.  (Compton cooling does not 
play an important role in our universe and will be considered in 
section~\ref{sec:comp}.)  In this section we review the conditions 
for cooling and derive the upper and lower bounds on the halo mass.

We stress that the present section is entirely about reviewing the 
conventional physics of cooling in {\em our\/} universe; ``fundamental'' 
parameters like $Q$ or $m_e$ are taken to have the observed values. 
In later sections, we will consider multiverse settings in which 
one or more parameters scan.  At that point we will make use of the 
parametric dependence of the expressions we derive here, in order 
to explain why certain parameters that scan over the multiverse 
are observed to take the values that they do.

We assume adiabatic, near scale-invariant, density perturbations with 
primordial amplitude $Q$.  Far into the matter dominated era, a density 
perturbation of mass $M$ grows linearly with the scale factor, $a(t)$, 
and at time $t$ has an amplitude
\begin{equation}
  Q(M,t) = \left(\frac{\rho(t)}{\rho_{\rm eq}}\right)^{1/3} f(M)\, Q,
\label{eq:Q}
\end{equation}
where $\rho(t)$ is the matter density and
\begin{equation}
  \rho_{\rm eq} = \frac{\pi^2 N_{\rm eq} T_{\rm eq}^4}{30}
\label{eq:rho-eq}
\end{equation}
is its value at the time $t_{\rm eq}$ when it equals the radiation 
density.  Here, $N_{\rm eq} \simeq 3.36$ is the number of relativistic 
species at the temperature of matter-radiation equality, and $T_{\rm eq}$ 
is the temperature at equality.  We will consider $T_{\rm eq}$ the 
fundamental parameter that defines the time of equality.

The function $f(M)$ includes both a mild logarithmic growth 
for perturbations that entered the horizon before $t_{\rm eq}$ 
and complicated behavior over a lengthy transition region 
between radiation and matter dominated eras.  For example, for 
$M = 10^{12} M_\odot$, which is close to the mass of the Milky Way, 
$f(M) \simeq 43$~\cite{Tegmark:2005dy}.

For sufficiently small $\rho_\Lambda$, the perturbation on scale $M$ 
goes non-linear and virializes at time defined by
\begin{equation}
  Q(M,t_{\rm vir}) = \delta_{\rm col} 
    \equiv \frac{3 (12 \pi)^{2/3}}{20} \simeq 1.69.
\label{eq:qmt}
\end{equation}
With Eq.~(\ref{eq:Q}), one finds
\begin{equation}
  t_{\rm vir} = c_{\rm vir} 
    \left( \frac{5\, \delta_{\rm col}^3}{\pi^3 N_{\rm eq}\, f(M)^3}\, 
    \frac{1}{G_{\rm N} \bar{\rho}}\right)^{1/2},
\label{eq:t_vir}
\end{equation}
where the coefficient $c_{\rm vir} \simeq 0.5$ is introduced to account 
for the effects of halo merging.%
\footnote{Halos grow by mergers and by accretion.  Both effects 
 contribute to the mass $M$ in Eq.~(\ref{eq:qmt}), and they cannot 
 be disentangled in a simple spherical collapse model.  However, 
 only major mergers lead to virialization.  In a more careful analysis, 
 the density and temperature of the virialized gas is set not at the 
 time when the halo reaches mass $M$, but at the earlier time when 
 it underwent its last major merger.  Our choice of $c_{\rm vir}$ 
 mimics this effect; we thank S.~Leichenauer for help with selecting 
 the value $0.5$.}
We have defined a fiducial density
\begin{equation}
  \bar{\rho} = Q^3 T_{\rm eq}^4,
\label{eq:rho_bar}
\end{equation}
which is closely related to the virial density of the halo
\begin{equation}
  \rho_{\rm vir} = \frac{f_\rho}{6\pi G_{\rm N} t_{\rm vir}^2} 
  = \frac{\pi^2 N_{\rm eq} f_\rho\, f(M)^3}{30\, \delta_{\rm col}^3\, 
    c_{\rm vir}^2}\, Q^3 T_{\rm eq}^4 
  \equiv \frac{\pi^2 N_{\rm eq} f_\rho\, f(M)^3}{30\, \delta_{\rm col}^3\, 
    c_{\rm vir}^2}\, \bar{\rho},
\label{eq:rho_vir}
\end{equation}
in an approximation where the virialized halo is taken to be spherically 
symmetric with uniform density.  Note that $\rho_{\rm vir}$ is a factor 
$f_\rho = 18\pi^2$ larger than the ambient density at $t_{\rm vir}$.

\subsection{Radiative cooling and galaxy formation}
\label{subsec:rad_cool}

Galaxies form in halos only if the virialized baryonic gas can cool 
and condense.  The temperature of the virialized halo is set by the 
gravitational potential energy gained by electrons and protons falling 
into the halo.  By the virial theorem,
\begin{equation}
  \frac{1}{2} \frac{3 G_{\rm N} M\mu}{5 R_{\rm vir}} 
  = \frac{3}{2}T_{\rm vir}.
\label{eq:vir-theor}
\end{equation}
We will take the average molecular mass $\mu$ to be $m_p/2$, so that
\begin{equation}
  T_{\rm vir} = \frac{G_{\rm N} \mu}{5} 
    \left(\frac{4\pi \rho_{\rm vir} M^2}{3}\right)^{1/3} 
  = \frac{\pi f(M)}{10\, \delta_{\rm col}} 
    \left( \frac{2 N_{\rm eq} f_\rho}{45\, c_{\rm vir}^2} \right)^{1/3} 
    G_{\rm N} m_p M^{2/3} \bar{\rho}^{1/3}.
\label{eq:T_vir}
\end{equation}
In our universe, $Q \simeq 2.0 \times 10^{-5}$ and $T_{\rm eq} 
\simeq 0.82~{\rm eV}$, so that a halo of mass $10^{12} M_\odot$ 
has $t_{\rm vir} \simeq 3.6 \times 10^9~{\rm years}$, $\rho_{\rm vir} 
\simeq (1.5 \times 10^{-2}~{\rm eV})^4$ and $T_{\rm vir} \simeq 
40~{\rm eV} \simeq 4.6 \times 10^5~{\rm K}$.

We can now state the conditions for cooling.  First, the temperature 
of the halo must be large enough to give significant ionization
\begin{equation}
  T_{\rm vir} > c_{\rm ion}\, \alpha^2 m_e,
\label{eq:T_vir-cond}
\end{equation}
where, for bremsstrahlung and hydrogen line cooling, $c_{\rm ion} 
\simeq 0.05$ so that $c_{\rm ion}\, \alpha^2 m_e \simeq 1.5 \times 
10^4~{\rm K}$. Using Eq.~(\ref{eq:T_vir}), this ionization requirement 
leads to a minimum value of the halo mass
\begin{equation}
  M \gtrsim M_- \equiv 
    \left( \frac{10\, \delta_{\rm col} c_{\rm ion}}{\pi f(M)} \right)^{3/2} 
    \left( \frac{45\, c_{\rm vir}^2}{2 N_{\rm eq} f_\rho} \right)^{1/2} 
    \frac{\alpha^3 m_e^{3/2}}{G_{\rm N}^{3/2} m_p^{3/2} \bar{\rho}^{1/2}}.
\label{eq:M_min}
\end{equation}
We will comment at the end of this section on the numerical value 
obtained here.

Second, the time scale for radiative cooling of the halo, 
$t_{\rm rad}$, should be less than the dynamical time scale 
$t_{\rm dyn}$ \cite{Rees:1977,Silk:1977wz,White:1978}, which 
we choose to be $t_{\rm vir}$:
\begin{equation}
  t_{\rm rad} < t_{\rm vir}.
\label{eq:t_rad}
\end{equation}
If $T_{\rm vir} \gtrsim 10^6\,{\rm K}$, the halo is fully ionized, and 
the radiative cooling rate is dominated by bremsstrahlung emission 
with a time scale at virialization of
\begin{equation}
  t_{\rm brems} = 
    \frac{9}{8K}\, \frac{m_e^{3/2} m_p T_{\rm vir}^{1/2}}{\alpha^3 \rho_B},
\label{eq:t_brems}
\end{equation}
where $K \simeq 3.17$, and $\rho_B = f_B\, \rho_{\rm vir}$ is the 
baryon density in the virialized halo with $f_B \simeq 1/6$.  At lower 
temperatures other processes are important, for example hydrogen line 
cooling for $10^4~{\rm K} \lesssim T \lesssim 10^6~{\rm K}$, so that 
we write the radiative cooling rate as
\begin{equation}
  t_{\rm rad} 
  = \left(\frac{9}{8K}\, f_{\rm rad}(T/m_e,\alpha) \right) 
    \frac{m_e^{3/2} m_p T_{\rm vir}^{1/2}}{\alpha^3 \rho_B}.
\label{eq:t_rad-2}
\end{equation}

The cooling condition of Eq.~(\ref{eq:t_rad}) is satisfied only by 
halos with mass
\begin{equation}
  M \lesssim M_+ \equiv 
  \left( \frac{2^7\, 5^{1/2}\, N_{\rm eq} f_\rho^{5/2} K^3 f_B^3\, 
      f(M)^3}{3^8\, \delta_{\rm col}^3 c_{\rm vir}^2 f_{\rm rad}^3 
    }\right) \frac{\alpha^9 \bar{\rho}}{G_{\rm N}^3 m_e^{9/2} m_p^{9/2}}.
\label{eq:M_max}
\end{equation}
We have thus translated the ionization and cooling rate conditions 
into lower and upper bounds on the halo mass, 
Eqs.~(\ref{eq:M_min},~\ref{eq:M_max}).

In our universe, $M_- \simeq 2.9 \times 10^9 M_\odot$ and $M_+ \simeq 
4.3 \times 10^{11} M_\odot$.  This interval is somewhat smaller 
than the observed range of halo masses; in particular, it does 
not include our own halo, with $M \sim 10^{12} M_\odot$.  This has 
a number of reasons.  We have worked with $1\sigma$ overdensities 
to associate a typical virial density and temperature to a given 
mass scale.  But a reasonable fraction of halos will form from, 
say, $2\sigma$ overdensities.  This can be incorporated into 
Eqs.~(\ref{eq:M_min},~\ref{eq:M_max}) by substituting $f(M) 
\to 2f(M)$; it lowers the minimum mass by about a factor of 
$3$ and raises the maximum mass by a factor of $8$.

With this definition, the range does include the Milky Way, if only 
barely.  This is as it should be: a halo of mass $10^{12} M_\odot$ is 
only just able to satisfy the cooling criteria~\cite{Tegmark:1997in}, 
corresponding to the coincidence that $t_{\rm cool} \sim t_{\rm vir}$. 
The lower bound, however, is still about two orders of magnitude 
higher than the smallest galaxies observed in our universe, which have 
$M \approx 10^7 M_\odot$.  This is because we have not included molecular 
cooling in our analysis. Molecular cooling can lower the cooling 
boundary by two orders of magnitude in our universe, but it will 
not have a significant effect in the neighborhood of the critical 
value of $\bar{\rho}$ which we will find in the following section. 
Hence we will ignore it here.

\subsection{Disruption of halo formation by a positive cosmological constant}
\label{subsec:halo-CC}

For definiteness, we will consider only positive values of the 
cosmological constant in this paper.  Negative values require a 
different analysis, though our ultimate conclusions would remain 
essentially unchanged.

A cosmological constant, $\Lambda>0$, or vacuum energy, $\rho_\Lambda 
=\Lambda/8\pi G_{\rm N}$, will come to dominate the evolution of the 
scale factor when the universe reaches an age of order $t_\Lambda 
= (3/\Lambda)^{1/2}$, when it begins to disrupt the formation of 
large scale structure.  At late times, it causes the universe to 
expand exponentially with characteristic timescale $t_\Lambda$.

A primordial overdensity which, in a flat universe with vanishing 
vacuum energy, would evolve into a halo with virial density 
$\rho_{\rm vir}$ will fail to evolve into a gravitationally bound 
object in a flat universe with~\cite{Weinberg:1987dv}
\begin{equation}
  \rho_\Lambda \gtrsim \frac{\rho_{\rm vir}}{54}.
\label{eq:weinberg}
\end{equation}
Using Eq.~(\ref{eq:rho_vir}), one finds that the observed vacuum 
energy,
\begin{equation}
  \rho_{\Lambda,o} = (2.3\times 10^{-3}~{\rm eV})^4
\label{eq:CC-observed}
\end{equation}
prevents the formation of halos of mass greater than $5.4 \times 
10^{14} M_\odot$ in our universe.

\subsection{Compton cooling}
\label{subsec:compton}

A proto-galaxy can also cool by inverse Compton scattering of electrons 
from the cosmic background radiation.  The cooling rate of a fully 
ionized plasma with temperature $T$ against a background with 
temperature $T_{\rm CMB} \ll T$ is given by
\begin{equation}
  \frac{\dot{T}}{T} = \frac{4 \sigma_{\rm T} a T_{\rm CMB}^4}{3 m_e},
\label{eq:comp-rate}
\end{equation}
where $\sigma_{\rm T} = 8\pi\alpha^2/3 m_e^2$ is the total Thompson 
cross section and $a = \pi^2/15$.  Note that this rate does not depend 
on the density or temperature of the plasma.  Its inverse defines the 
Compton cooling time scale,
\begin{equation}
  t_{\rm comp} = \frac{135 m_e^3}{32\pi^3 \alpha^2 T_{\rm CMB}^4},
\label{eq:t_comp}
\end{equation}
which depends on cosmological time $t$ through the CMB temperature
\begin{equation}
  T_{\rm CMB}(t) = \frac{5}{\pi^4 N_{\rm eq} G_{\rm N} T_{\rm eq} t^2}.
\label{eq:T_CMB}
\end{equation}

Compton cooling is effective as long as this time scale, at the time 
of virialization, $t_{\rm vir}$, is faster than the dynamical time 
scale of the virialized halo, which is also given by $t_{\rm vir}$:
\begin{equation}
  t_{\rm comp}(t_{\rm vir}) < t_{\rm vir}.
\label{eq:Comp-cond}
\end{equation}
Thus, Compton cooling is effective for any halos that virialize prior 
to the time
\begin{equation}
  t_{\rm comp,max} \simeq 
    \frac{5^{1/5}\, 2^3}{3^{9/5}\pi^{7/5} N_{\rm eq}^{4/5}}\, 
    \frac{\alpha^{6/5}}{G_{\rm N}^{4/5} T_{\rm eq}^{4/5} m_e^{9/5}}.
\label{eq:tcompmax}
\end{equation}
In our universe, this time is approximately $0.36~{\rm Gyr}$.

Even before this time, Compton cooling may not be the fastest 
cooling process.  It will be faster than radiative cooling only 
if $t_{\rm comp} < t_{\rm rad}$.  However, we shall not need 
to investigate this condition separately in this paper, because we 
will be interested in situations where Compton cooling is effective 
for halos in which radiative cooling is on the verge of being 
ineffective, i.e., for halos with $t_{\rm vir} \sim t_{\rm rad}$.

\section{Predicting the Virial Density and Mass of Galactic Halos}
\label{sec:pred_cool}

In this section, we will consider the consequences of allowing the 
characteristic virial density, $\bar{\rho}$, to scan in the multiverse. 
In section~\ref{subsec:boundary}, we use the parametric expressions for 
the two cooling boundaries on the halo mass $M$ derived in the previous 
section to show that there is a catastrophic lower bound on $\bar{\rho}$, 
where the maximum and minimum mass become equal.  We assume a multiverse 
force that pushes $\bar{\rho}$ towards the catastrophic galaxy cooling 
boundary.  In section~\ref{subsec:basic_pred}, we use this qualitative 
assumption to obtain rough quantitative predictions of the mass of 
a typical halo and its virial density, in terms of $\alpha$, $m_e$, 
$m_p$, and $G_{\rm N}$.  In section~\ref{subsec:multi-force}, we compare 
our result to astrophysical data.  As expected, our universe is near 
but not exactly at the catastrophic boundary.  The distance to the 
boundary allows us to derive empirical bounds on the strength of 
the multiverse force.

\subsection{The catastrophic cooling boundary}
\label{subsec:boundary}

Both the upper and lower bound on halo masses, 
Eqs.~(\ref{eq:M_min},~\ref{eq:M_max}), depend on the primordial 
density contrast and the temperature at equality only through the 
combination $\bar{\rho} = Q^3 T_{\rm eq}^4$, which can be thought of 
as the characteristic virial density in the universe.  We will now 
assume that $\bar{\rho}$ is a variable that can take on different 
values in the multiverse.  The origin of this scanning can be via $Q$ 
or $T_{\rm eq}$, but we need not specify which in the analysis here. 
For now, we will regard Eqs.~(\ref{eq:M_min},~\ref{eq:M_max}) as 
functions of the scanning parameter $\bar{\rho}$, and we will phrase 
our assumptions in terms of the multiverse distribution function 
for $\bar{\rho}$.  (Later, in section~\ref{sec:comp}, we will 
consider an additional boundary that distinguishes between $Q$ 
and $T_{\rm eq}$.)

\begin{figure}[t]
\begin{center}
\begin{picture}(390,185)(11,-36)
%
%
  \Text(80,-23)[t]{\large (a)}
  \Line(-20,0)(180,0) \Line(180,0)(173,4) \Line(180,0)(173,-4)
  \Line(-10,-10)(-10,140) \Line(-10,140)(-14,133) \Line(-10,140)(-6,133)
  \Text(179,-8)[t]{\large $\bar{\rho}$} \Text(-16,139)[r]{\large $M$}
  \Line(10,50)(150,140) \Line(10,70)(180,10)
  \Text(15,126)[l]{no radiative} \Text(30,114)[l]{cooling}
  \Text(50,25)[l]{no} \Text(55,15)[l]{ionization}
  \Vertex(44.5,69.1){2.5}
  \Line(11.47,69.47)(10.00,68.00)
  \Line(15.17,68.17)(10.00,63.00)
  \Line(18.86,66.86)(12.86,60.86)
  \Line(10.0,53.0)(26.6,69.6)
  \Line(15.6,53.6)(40.6,78.6)
  \Line(23.95,56.95)(54.6,87.6)
  \Line(43.6,71.6)(68.6,96.6)
  \Line(57.6,80.6)(82.6,105.6)
  \Line(71.6,89.6)(96.6,114.6)
  \Line(85.6,98.6)(110.6,123.6)
  \Line(99.6,107.6)(124.6,132.6)
  \Line(113.6,116.6)(137.0,140.0)
  \Line(127.6,125.6)(142.0,140.0)
  \Line(141.6,134.6)(147.0,140.0)
  \Line(33.65,61.65)(27.65,55.65)
  \Line(37.34,60.34)(31.34,54.34)
  \Line(41.04,59.04)(35.04,53.04)
  \Line(44.73,57.73)(38.73,51.73)
  \Line(48.43,56.43)(42.43,50.43)
  \Line(52.13,55.13)(46.13,49.13)
  \Line(55.82,53.82)(49.82,47.82)
  \Line(59.52,52.52)(53.52,46.52)
  \Line(63.21,51.21)(57.21,45.21)
  \Line(66.91,49.91)(60.91,43.91)
  \Line(70.60,48.60)(64.60,42.60)
  \Line(74.30,47.30)(68.30,41.30)
  \Line(78.00,46.00)(72.00,40.00)
  \Line(81.69,44.69)(75.69,38.69)
  \Line(85.39,43.39)(79.39,37.39)
  \Line(89.08,42.08)(83.08,36.08)
  \Line(92.78,40.78)(86.78,34.78)
  \Line(96.47,39.47)(90.47,33.47)
  \Line(100.17,38.17)(94.17,32.17)
  \Line(103.86,36.86)(97.86,30.86)
  \Line(107.56,35.56)(101.56,29.56)
  \Line(111.26,34.26)(105.26,28.26)
  \Line(114.95,32.95)(108.95,26.95)
  \Line(118.65,31.65)(112.65,25.65)
  \Line(122.34,30.34)(116.34,24.34)
  \Line(126.04,29.04)(120.04,23.04)
  \Line(129.73,27.73)(123.73,21.73)
  \Line(133.43,26.43)(127.43,20.43)
  \Line(137.13,25.13)(131.13,19.13)
  \Line(140.82,23.82)(134.82,17.82)
  \Line(144.52,22.52)(138.52,16.52)
  \Line(148.21,21.21)(142.21,15.21)
  \Line(151.91,19.91)(145.91,13.91)
  \Line(155.60,18.60)(149.60,12.60)
  \Line(159.30,17.30)(153.30,11.30)
  \Line(163.00,16.00)(157.00,10.00)
  \Line(166.69,14.69)(160.69,8.69)
  \Line(170.39,13.39)(164.39,7.39)
  \Line(174.08,12.08)(168.08,6.08)
  \Line(177.78,10.78)(171.78,4.78)
  \Line(180.00,8.00)(175.47,3.47)
  \Line(180.00,3.00)(179.17,2.17)
  \Line(119.8,68.1)(100.0,68.1) \Line(119.8,67.9)(100.0,67.9)
  \Line(100,68)(107,71) \Line(100,68)(107,65)
  \Text(128,69)[l]{\large ${\bf F}$}
  \DashLine(30.08,62.91)(30.08,0){2}
  \Text(31,-5)[t]{\large $\bar{\rho}_{\rm c}$}
  \DashLine(30.08,62.91)(-10,62.91){2}
  \Text(-13,63)[r]{\large $M_{\rm c}$}
%
%
  \Text(330,-23)[t]{\large (b)}
  \Line(230,0)(430,0) \Line(430,0)(423,4) \Line(430,0)(423,-4)
  \Line(240,-10)(240,140) \Line(240,140)(236,133) \Line(240,140)(244,133)
  \Text(429,-8)[t]{\large $\bar{\rho}$} \Text(234,139)[r]{\large $M$}
  \Line(260,50)(400,140) \Line(260,70)(430,10)
  \Vertex(294.5,69.1){2.5}
  \DashCArc(264.0,69.5)(0.8,80,260){1}
  \DashLine(265.1,70.2)(277.0,67.4){1}
  \DashCArc(277.8,67.9)(0.6,280,65){1}
  \DashLine(277.2,68.9)(274.8,70.1){1}
  \DashCArc(274.2,71.1)(0.7,80,255){1}
  \DashLine(274.9,71.7)(283.9,68.7){1}
  \DashCArc(286.4,73.4)(5.2,251.6,302.7){1}
  \Line(260.5,82.1)(267.0,70.8)
  \Line(267.0,70.8)(264.4,72.3) \Line(267.0,70.8)(267.0,73.8)
  \Text(247,96.0)[l]{\small line} \Text(252,87.0)[l]{\small cooling}
  \DashCArc(290.8,61.6)(6.8,180.0,250.6){1}
  \DashLine(289.6,54.8)(430.0,5.3){1}
  \Line(278.0,45.4)(285.5,56.0)
  \Line(285.5,56.0)(285.5,53.0) \Line(285.5,56.0)(282.9,54.5)
  \Text(247,40.5)[l]{\small molecular} \Text(252,30.5)[l]{\small cooling}
  \DashLine(299,130)(299,10){4} 
  \DashLine(314,130)(314,10){4} 
  \DashLine(329,130)(329,10){4} 
  \Line(283.7,119.6)(297.5,108.0)
  \Line(297.5,108.0)(293.6,108.7) \Line(297.5,108.0)(296.1,111.8)
  \Text(247,136.5)[l]{\small Compton} \Text(252,126.5)[l]{\small cooling}
  \Line(303,128)(345,128)
  \Line(345,128)(339,130)\Line(345,128)(339,126)
  \Text(332,134)[bl]{\tiny $Q\!\!\searrow$ or $T_{\rm eq}\!\!\nearrow$}
\end{picture}
\caption{(a) Simplified sketch of the boundaries for ionization (lower) 
 and radiative cooling (upper) for a halo of mass $M$.  Our galaxy, 
 shown by the dot, is very close to the radiative cooling boundary 
 and has a virial density only $1~\mbox{--}~2$ orders of magnitude 
 larger than the minimum possible.  (b) Dotted lines show modifications 
 to the ionization and radiative cooling boundaries from molecular 
 cooling and atomic line cooling.  To the right of the dashed line, 
 halos are able to cool by inverse Compton scattering from the 
 background radiation.}
\label{fig:gal}
\end{center}
\end{figure}
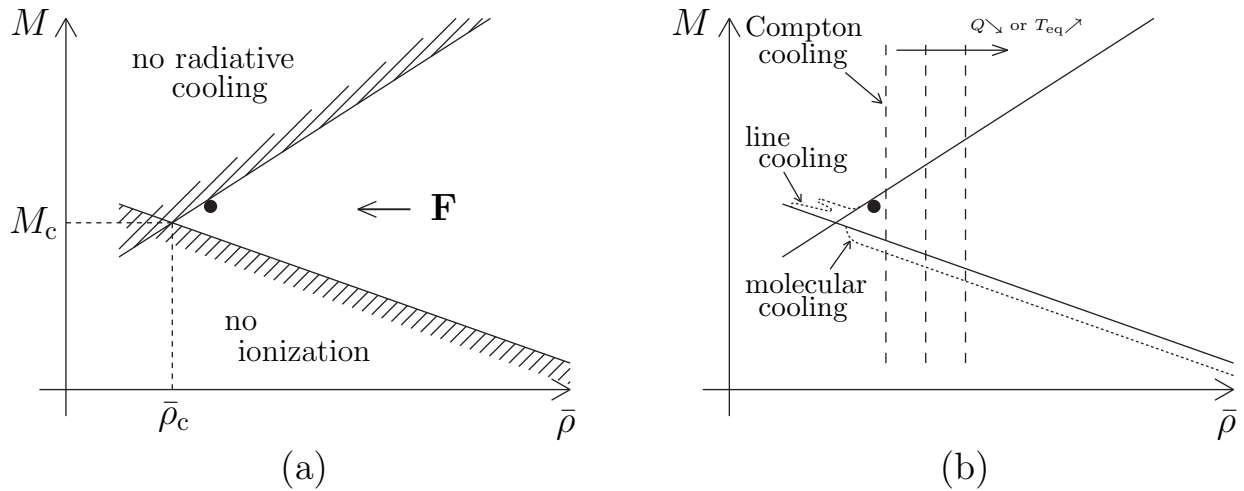

These functions are sketched in Fig.~\ref{fig:gal}a (with additional 
cooling boundaries given in Fig.~\ref{fig:gal}b).
Apart from the mild logarithmic factor $f(M)$, they are both simple 
power laws, one increasing and one decreasing with $\bar{\rho}$.  They 
cross at the critical value
\begin{equation}
  \bar{\rho}_{\rm c} = \left( 
    \frac{3^6\times 5\, \delta_{\rm col}^3\, c_{\rm ion}\, c_{\rm vir}^2 
    f_{\rm rad}^2}{2^4\pi f(M)^3 N_{\rm eq} f_\rho^2 K^2 f_{\rm B}^2} 
    \right) \frac{G_{\rm N} m_e^4 m_p^2}{\alpha^4} 
  \simeq (0.97\times 10^{-4}~{\rm eV})^4.
\label{eq:rho_c}
\end{equation}
It is important to note that $M$ is not a scanning parameter.  For any 
value of $\bar{\rho}$, halos with all values of $M$ will virialize.%
\footnote{This assumes that curvature and vacuum energy are negligible. 
 We will consider nonzero vacuum energy in later sections.}
We will assume, for simplicity, that observers form as long as there 
exists {\em some\/} range of $M$, no matter how small, that can 
cool.  Thus, $\bar{\rho}_{\rm c}$ represents a catastrophic boundary 
in $\bar{\rho}$ space.  Universes with $\bar{\rho} > \bar{\rho}_{\rm c}$ 
contain observers; universes with $\bar{\rho} < \bar{\rho}_{\rm c}$ 
do not.

At the catastrophic boundary, only halos with critical mass
\begin{equation}
  M_{\rm c} = M_+(\bar{\rho}_{\rm c}) = M_-(\bar{\rho}_{\rm c}) \simeq 
    \Biggl( \frac{40\sqrt{5}\, f_\rho^{1/2} c_{\rm ion}\, K f_B}{9\pi\, 
    f_{\rm rad}}\Biggr) \frac{\alpha^5}{G_{\rm N}^2 m_e^{1/2} m_p^{5/2}} 
  \simeq 2.4 \times 10^{10} M_\odot
\label{eq:M_c}
\end{equation}
and critical virial density
\begin{equation}
  \rho_{\rm vir,c} \simeq 
    \Biggl( \frac{3^5 \pi\, c_{\rm ion} f_{\rm rad}^2}{2^5 f_\rho 
    K^2 f_B^2} \Biggr) \frac{G_{\rm N} m_e^4 m_p^2}{\alpha^4} 
  \simeq (7.7 \times 10^{-3}~{\rm eV})^4
\label{eq:rho_vir_c}
\end{equation}
are able to cool.

Although we have been keeping track of numerical factors in the above 
equations, we aim only to define the position of this catastrophic 
boundary at the order of magnitude level.  For example, there is 
considerable uncertainty in the numerical factor in the inequality of 
Eq.~(\ref{eq:t_rad}).  The physics of cooling involves non-uniformities 
and shocks and, furthermore, the boundary is not completely sharp, 
with the statistics of merging playing an important role.  Moreover, 
as discussed in the previous section, there is some ambiguity in how 
$M_+$ and $M_-$ should be defined.  These ambiguities arise in part 
because we are treating the catastrophic boundary as sharp; in a 
more detailed analysis, the anthropic weighting factor would change 
continuously over a range of values of $\bar{\rho}$.  Even then, 
a precise estimate of the number of observers would remain a challenge.

\subsection{Predictions}
\label{subsec:basic_pred}

To obtain a prediction, we assume that the multiverse distribution 
$f(\bar{\rho})$ is a decreasing function.  Thus, multiverse statistics 
favors {\em low} values of $\bar{\rho}$, as shown by the direction of 
the arrow in Fig.~\ref{fig:gal}a.  In this case, observers typically 
measure a value of $\bar{\rho}$ close to the critical value:
\begin{equation}
  \bar{\rho} \sim \bar{\rho}_{\rm c}.
\label{eq:rho-bar_pred}
\end{equation}
Most universes have lower values of $\bar{\rho}$, so that no 
perturbations are able to cool; these universes will not be observed. 
Observers live only in universes with $\bar{\rho}$ sufficiently large 
to allow cooling.  Most will find themselves near the catastrophic 
boundary, with $\bar{\rho}$ larger, but not much larger, than 
$\bar{\rho}_{\rm c}$ given in Eq.~(\ref{eq:rho_vir_c}).

There are a number of corollaries to this prediction.  Galaxies 
in these universes will arise in a relatively narrow range of halo 
masses, controlled by the critical value $M_{\rm c}$ defined in 
Eq.~(\ref{eq:M_c}), and ranging from a minimum halo mass of order 
$M_- \approx M_{\rm c}(\bar{\rho}/\bar{\rho}_{\rm c})^{-1/2}$ to a maximum 
halo mass of order $M_+ \approx M_{\rm c}(\bar{\rho}/\bar{\rho}_{\rm c})$. 
(The minimum mass can be lower with molecular cooling, and we have 
neglected logarithmic factors.)  In these halos, virial densities 
will be larger, but not much larger, than the critical value given 
in Eq.~(\ref{eq:rho_vir_c}); and the timescale for radiative galaxy 
cooling will not be much shorter than the timescale for virialization.

For the sake of clarity, let us summarize these results by restating 
the relevant formulas without logarithmic or order one factors.  By 
allowing $Q$ and/or $T_{\rm eq}$ to scan, and assuming a multiverse 
force towards small values of $\bar{\rho} = Q^3 T_{\rm eq}^4$, we were 
able to predict the observed virial density at the order of magnitude 
level from fundamental parameters:
\begin{equation}
  \rho_{\rm vir} \sim \frac{G_{\rm N} m_e^4 m_p^2}{\alpha^4}.
\label{eq:rho_vir-pred}
\end{equation}
It followed that our universe should be close to the radiative cooling 
boundary of Eq.~(\ref{eq:t_rad}):
\begin{equation}
  t_{\rm rad} \sim t_{\rm vir} \sim \frac{\alpha^2}{G_{\rm N} m_e^2 m_p},
\label{eq:tcool=tvir}
\end{equation}
explaining one of the coincidences of Eq.~(\ref{eq:coscoinc}); and 
that the galactic mass scale arises from%
\footnote{Arguments that radiative cooling introduces a critical mass 
 scale for galaxies, given by Eq.~(\ref{eq:M_c}), were made in the late 
 1970s \cite{Rees:1977,Silk:1977wz,White:1978,Carr:1979sg}. Since a 
 large range of halos on one side of the boundary can cool, however, 
 a multiverse force and anthropic arguments are needed to explain 
 why observed galaxies are so close to this critical value.}
\begin{equation}
  M \sim \frac{\alpha^5}{G_{\rm N}^2 m_e^{1/2} m_p^{5/2}}.
\label{eq:M-pred}
\end{equation}
Another corollary is worth pointing out.  The radiative cooling time 
scale at the critical point, Eq.~(\ref{eq:t_rad}), has precisely the 
same parametric form as the lifetime for a main sequence hydrogen 
burning star of maximum mass, $t_{\rm burn}$.  Hence, in a typical 
universe that is not far from the critical point, there is no freedom 
to choose the stellar lifetime: $t_{\rm burn} \sim t_{\rm rad} \sim 
t_{\rm vir}$.

These predictions are successful.  In our universe, $\bar{\rho}_o = 
Q_o^3 T_{{\rm eq},o}^4 \simeq (2.4\times 10^{-4}~{\rm eV})^4$, which 
is only a factor of $40$ larger than the critical value:
\begin{equation}
  \bar{\rho}_o = 40\, \bar{\rho}_{\rm c}.
\label{eq:rho_o}
\end{equation}
(We have restored order one factors.)  This is remarkable since these 
two quantities could have been many orders of magnitude apart.  It 
would appear even more impressive had we compared quantities of mass 
dimension one: $\bar{\rho}_o^{1/4} = 2.5\, \bar{\rho}_{\rm c}^{1/4}$.
Nevertheless, $\bar{\rho}_o$ is sufficiently larger than 
$\bar{\rho}_{\rm c}$ that in our universe galaxies with 
a considerable range of mass can cool radiatively.  Our 
galaxy is near the upper end of this range, suggesting further 
environmental selection in our universe.

\subsection{The strength of the multiverse force}
\label{subsec:multi-force}

The strength of the multiverse force determines how close a typical 
universe is to the critical point.  For $\bar{\rho}$, this strength 
cannot be predicted from first principles with current theoretical 
technology.  Instead, let us now reverse course and put observational 
constraints on the force.  From Eqs.~(\ref{eq:force-p},~\ref{eq:rho_o}), 
we conclude that the force satisfies
\begin{equation}
  -p_{\bar{\rho}} = 
    0.19~^{+0.31}_{-0.14}(1\sigma)~^{+0.84}_{-0.18}(2\sigma).
\label{eq:force_rho-bar}
\end{equation}

Note that the force discussed here is the effective force obtained after 
integrating out parameters other than $\bar{\rho}$~\cite{Hall:2007ja}, 
such as the cosmological constant.  In the following sections we will 
consider such parameters, and we will be able to determine under which 
conditions a force in the above range can arise.  We will see that this 
places interesting constraints on the multiverse distribution function, 
and in particular, on the measure for regulating the infinities of 
eternal inflation.

\section{Predicting the Cosmological Constant}
\label{sec:CC-1}

In this section, we will allow a second parameter to scan: the 
cosmological constant or vacuum energy, $\rho_\Lambda$.%
\footnote{For definiteness, we will consider only positive values 
 of the cosmological constant, although it is not difficult to extend 
 our analysis to negative values.}
To avoid runaway, we need a second boundary.  One way to obtain such 
a boundary is to require, as in the previous section, that galaxies 
form~\cite{Weinberg:1987dv}.  This condition is clearly necessary; 
whether it is the most important constraint on $\rho_\Lambda$ depends 
on the choice of measure.  As shown in Ref.~\cite{Bousso:2008hz}, 
the version of the scale factor measure formulated in 
Ref.~\cite{DeSimone:2008bq} leads to this constraint.

We will be able to explain the coincidence $t_{\rm vir} \sim t_\Lambda$ 
in this way.  We will also be able to use observed values of $\bar{\rho}$ 
and $\rho_\Lambda$ to constrain the two-dimensional multiverse forces 
on these parameters.  We will find, however, that these empirical 
constraints lead to tension with theoretical expectations.  This, 
therefore, disfavors the measure we consider here.

In the next section we will consider a different class of measures 
which will be more successful, and also more powerful in that they 
constrain an additional scanning parameter.

\subsection{Catastrophic boundary from large scale structure}
\label{subsec:cat-LSS}

\paragraph{The role of the measure}
In theories with at least one long-lived metastable vacuum with
positive cosmological constant, a finite false vacuum region can
evolve into an infinite four-volume containing infinitely many bubbles
of other vacua.  Each bubble is an infinite open universe; if it
contains any observers, no matter how sparse, it will contain
infinitely many.  These infinities must be regulated if we are to say
that some outcomes of observations are typical and others unlikely.
This is the measure problem of eternal inflation.

We will consider four measure proposals that are reasonably 
well-defined and do not suffer from known catastrophic problems.%
\footnote{The measures used in the older literature on environmental 
 selection of the cosmological constant, such as observers-per-baryon, 
 fail catastrophically in eternally inflating universes driven by 
 false vacua \cite{Page:2006dt,Bousso:2006xc}, so they will not 
 be considered here.  For other examples of catastrophic problems, 
 see, e.g., Ref.~\cite{Bousso:2007nd}.}
The first and second are the causal patch~\cite{Bousso:2006ev} (CP) 
and causal diamond (CD)~\cite{Bousso:2007kq} cutoffs; the third and 
fourth are versions of the scale factor cutoff, which we will refer 
to as SF1~\cite{DeSimone:2008bq} and SF2~\cite{Bousso:2008hz}.  We 
will not review these measures here in detail, but focus instead 
on their implications for the analysis of the cosmological constant 
problem.  The measures differ in important ways: they lead not 
only to different catastrophic boundaries, but also to different 
multiverse forces on the cosmological constant and virial density.

Both measures are based on a geometric cutoff: one computes the 
expected number of observations made in some finite spacetime region. 
They differ in how that region is defined.  The causal patch measure 
(CP) restricts to the causal past of the future endpoint of a geodesic 
in the spacetime.  In long-lived metastable vacua with positive 
cosmological constant, this definition reduces to counting observations 
that take place within the cosmological (de~Sitter) event horizon 
surrounding the geodesic.  This has two important consequences. 
First, observations made after vacuum domination ($t_{\rm obs} 
> t_\Lambda$) contribute negligibly; they are exponentially suppressed 
by the accelerated expansion that empties the horizon region of 
matter.  This leads to a catastrophic boundary $\Lambda \lesssim 
t_{\rm obs}^{-2}$.%
\footnote{This boundary is stronger than the catastrophic boundary 
 coming from the disruption of galaxy formation.  Of course, nothing 
 prevents the formation of observers in a universe where suitable 
 galaxies form.  However, such observers are not counted if the 
 galaxies are not contained in the cutoff region.  The existence 
 of such a boundary, coming entirely from the measure, may seem 
 counter-intuitive; indeed, it is among the most interesting lessons 
 learned from recent studies of the measure problem~\cite{Bousso:2007kq}. 
 In eternal inflation, no conventional anthropic boundary is 
 well-defined without a measure.  For example, with $\Lambda \gg 
 t_{\rm vir}^{-2}$, there is still a tiny amplitude for the formation 
 of galaxies from unusually strong density perturbations, so there 
 will be infinitely many observers no matter whether $\Lambda$ is 
 above or below the Weinberg bound.  To make a meaningful comparison, 
 we need to pick a measure.  The measures usually considered are 
 geometric cutoffs that necessarily depend on quantities that affect 
 the geometry, such as $\Lambda$.  It should not surprise us, then, 
 that they can introduce catastrophic boundaries on observable 
 parameters.}
Secondly, at all times before vacuum domination ($t < t_\Lambda$), 
the total matter mass inside the causal patch scales as 
$\Lambda^{-1/2}$.  For smaller $\Lambda$, more matter is included, 
which mitigates the landscape force towards large values of $\Lambda$.

The causal diamond (CD) cutoff further restricts the surviving 
spacetime region.  A point is included only if it lies both in the 
causal patch and in the causal future of the intersection of the 
generating geodesic with the reheating hypersurface.  This leads to 
the same catastrophic boundary, $\Lambda \lesssim t_{\rm obs}^{-2}$. 
However, before $t_\Lambda$, the comoving volume of the diamond is 
controlled by the future light-cone.  Thus it is independent of 
$\Lambda$ and grows linearly with time.  This favors later observers.

As shown in Ref.~\cite{Bousso:2008hz}, the scale factor cutoff (SF1) 
can be defined as a small neighborhood, of fixed physical volume, 
of a geodesic in the multiverse.  (The formulation of the scale factor 
measure given in Ref.~\cite{DeSimone:2008bq} is not completely 
equivalent but shares the following properties~\cite{Bousso:2008hz}%
\footnote{This was not originally recognized in 
 Ref.~\cite{DeSimone:2008bq}, who concluded that the properties 
 of their formulation were similar to those ascribed here to SF2.}
relevant for our analysis.)  Like in the CD measure, and unlike CP, 
the size of the SF1 cutoff region is independent of $\Lambda$ before 
the time $t_\Lambda$.  Because the geodesic is likely to become 
trapped in a collapsed region, and one is including only its immediate 
neighborhood, the expansion of the universe after $t_{\rm vir}$ does 
not reduce the number of observers contained in the cutoff region. 
This leads to two important differences from both CD and CP.  First, 
there is no suppression for $t_{\rm obs} > t_\Lambda$, so the only 
relevant catastrophic boundary is the Weinberg bound, $\Lambda 
\lesssim t_{\rm vir}^{-2}$.  And secondly, in SF1 the number of 
observers is proportional to their physical density near the geodesic; 
this favors larger virial density.

SF2 is a hybrid of SF1 and CD.  This measure has only been tentatively 
formulated~\cite{Bousso:2008hz}, but we include it here because we are 
mainly interested in the consequences of a measure with the following 
properties: like in the CD measure, the catastrophic boundary is 
$\Lambda \lesssim t_{\rm obs}^{-2}$, and the spatial size of the 
cutoff region is $\Lambda$-independent before $t_\Lambda$.  The 
number of observers is proportional to the average density of matter 
in the universe at the time when observations are made.  This is 
similar to SF1, except that for SF1 the relevant density is set at 
an earlier time, when the first collapsed structures form.

In this section, we begin by considering the catastrophic boundary 
relevant to SF1: the disruption of the formation of galactic halos. 
In section~\ref{sec:CC-2}, we will consider the catastrophic boundary 
relevant to CP, CD, and SF2: the exponential dilution of the number 
density of galaxies.  For each case, we will obtain predictions for 
scanning parameters and empirical constraints on multiverse forces. 
We will discuss whether these constraints are compatible with 
theoretical expectations, which also depend on the measure.

\paragraph{The boundary}
In a given universe, the cosmological constant disrupts the formation 
of halos above a certain mass, because the logarithmic factor 
$f(M)$ in Eq.~(\ref{eq:rho_vir}) will enter into the condition 
of Eq.~(\ref{eq:weinberg}).  It is possible that there is a minimum 
galaxy mass necessary for the formation of observers.  As in the 
previous section, however, we will work with the conservative 
assumption that any galaxies will do.  Thus, a catastrophic boundary 
for the cosmological constant is reached only when $\rho_\Lambda$ is 
so large as to disrupt the formation of the smallest halos that can 
cool.  By Eq.~(\ref{eq:weinberg}), this corresponds to the condition
\begin{equation}
  \rho_{\Lambda,{\rm max}}(\bar{\rho})
  = \frac{\rho_{\rm vir}(M_-(\bar{\rho}))}{54},
\label{eq:rlm}
\end{equation}
where $\rho_{\rm vir}(M_-)$ and $M_-(\bar{\rho})$ are given in 
Eqs.~(\ref{eq:rho_vir}) and (\ref{eq:M_min}).

\begin{figure}[t]
\begin{center}
\begin{picture}(395,157)(6,-20)
%
%
  \Line(-25,0)(155,0) \Line(155,0)(148,4) \Line(155,0)(148,-4)
  \Line(-15,-10)(-15,130) \Line(-15,130)(-19,123) \Line(-15,130)(-11,123)
  \Text(-21,128)[r]{\large $\rho_\Lambda$}
  \Text(153,-9)[t]{\large $\bar{\rho}$}
  \Line(-5,10)(115,130)
  \Line(-3,12)(-3,23.3)
  \Line(2,17)(2,28.3)
  \Line(7,22)(7,33.3)
  \Line(12,27)(12,38.3)
  \Line(17,32)(17,43.3)
  \Line(22,37)(22,48.3)
  \Line(27,42)(27,53.3)
  \Line(32,47)(32,58.3)
  \Line(37,52)(37,63.3)
  \Line(42,57)(42,68.3)
  \Line(47,62)(47,73.3)
  \Line(52,67)(52,78.3)
  \Line(57,72)(57,83.3)
  \Line(62,77)(62,88.3)
  \Line(67,82)(67,93.3)
  \Line(72,87)(72,98.3)
  \Line(77,92)(77,103.3)
  \Line(82,97)(82,108.3)
  \Line(87,102)(87,113.3)
  \Line(92,107)(92,118.3)
  \Line(97,112)(97,123.3)
  \Line(102,117)(102,128.3)
  \Line(107,122)(107,130)
  \Line(112,127)(112,130)
  \Line(64.9,25.0)(64.9,54.8) \Line(65.1,25.0)(65.1,54.8)
  \Line(65,55)(62,48) \Line(65,55)(68,48)
  \Text(70,25)[l]{\large ${\bf F}$}
  \DashLine(69,71)(91,93){2}
  \Line(91,93)(88.46,87.56) \Line(91,93)(85.56,90.46)
%
%
  \Line(190,63)(215,63) \Line(210,66)(220,60)
  \Line(190,57)(215,57) \Line(210,54)(220,60)
  \Line(192,63)(190,61) \Line(194,63)(190,59) \Line(196,63)(190,57)
  \Line(198,63)(192,57) \Line(200,63)(194,57) \Line(202,63)(196,57)
  \Line(204,63)(198,57) \Line(206,63)(200,57) \Line(208,63)(202,57)
  \Line(210,63)(204,57) \Line(212,63)(206,57) \Line(214,63)(208,57)
  \Line(215.7,62.7)(210,57) \Line(217.0,62.0)(212,57)
  \Line(218.2,61.2)(214,57) \Line(219.4,60.4)(217.6,58.6)
  \Text(205,47)[t]{$t_\Lambda \sim t_{\rm vir}$}
%
%
  \Line(250,0)(430,0) \Line(430,0)(423,4) \Line(430,0)(423,-4)
  \Line(260,-10)(260,130) \Line(260,130)(256,123) \Line(260,130)(264,123)
  \Text(254,128)[r]{\large $\rho_\Lambda$}
  \Text(428,-9)[t]{\large $\bar{\rho}$}
  \Line(270,10)(390,130)
  \Line(272,12)(272,23.3)
  \Line(277,17)(277,28.3)
  \Line(282,22)(282,33.3)
  \Line(287,27)(287,38.3)
  \Line(292,32)(292,43.3)
  \Line(297,37)(297,48.3)
  \Line(302,42)(302,53.3)
  \Line(307,47)(307,58.3)
  \Line(312,52)(312,63.3)
  \Line(317,57)(317,68.3)
  \Line(322,62)(322,73.3)
  \Line(327,67)(327,78.3)
  \Line(332,72)(332,83.3)
  \Line(337,77)(337,88.3)
  \Line(342,82)(342,93.3)
  \Line(347,87)(347,98.3)
  \Line(352,92)(352,103.3)
  \Line(357,97)(357,108.3)
  \Line(362,102)(362,113.3)
  \Line(367,107)(367,118.3)
  \Line(372,112)(372,123.3)
  \Line(377,117)(377,128.3)
  \Line(382,122)(382,130)
  \Line(387,127)(387,130)
  \Line(315,-7)(315,130)
  \Text(317,-12)[t]{\large $\bar{\rho}_{\rm c}$}
  \Line(307,127)(310,130)
  \Line(307,120)(315,128)
  \Line(307,113)(315,121)
  \Line(307,106)(315,114)
  \Line(307,99)(315,107)
  \Line(307,92)(315,100)
  \Line(307,85)(315,93)
  \Line(307,78)(315,86)
  \Line(307,71)(315,79)
  \Line(307,64)(315,72)
  \Line(307,57)(315,65)
  \Line(307,50)(315,58)
  \Line(307,43)(315,51)
  \Line(307,36)(315,44)
  \Line(307,29)(315,37)
  \Line(307,22)(315,30)
  \Line(307,15)(315,23)
  \Line(307,8)(315,16)
  \Line(307,1)(315,9)
  \Line(307,-6)(315,2)
  \Line(313,-7)(315,-5)
  \Vertex(325,54){2.5}
  \Line(400.017,65.098)(370.670,70.273)
  \Line(399.983,64.902)(370.636,70.077)
  \Line(370.45,70.21)(377.865,71.95)
  \Line(370.45,70.21)(376.823,66.04)
  \Text(400,76)[l]{\large ${\bf F}$}
  \Line(400.017,40.098)(370.670,45.273)
  \Line(399.983,39.902)(370.636,45.077)
  \Line(370.45,45.21)(377.865,46.95)
  \Line(370.45,45.21)(376.823,41.04)
  \Line(400.017,15.098)(370.670,20.273)
  \Line(399.983,14.902)(370.636,20.077)
  \Line(370.45,20.21)(377.865,21.95)
  \Line(370.45,20.21)(376.823,16.04)
\end{picture}
\caption{When both $\rho_\Lambda$ and the virial density $\bar{\rho}$ 
 scan, the resulting behavior depends on the direction of the multiverse 
 force and on the boundaries.  On the left, a multiverse force with 
 only a $\rho_\Lambda$ component leads to runaway behavior along the 
 virialization boundary, as shown by the dashed arrow.  On the right, 
 a component of the force in the direction of $\bar{\rho}$ is added, 
 as well as the boundary from galactic radiative cooling.  Typical 
 universes lie at the tip of the cone, indicated by the dot.  Strong 
 forces would locate the dot within a factor of two of each boundary. 
 But the forces may be mild, placing the dot one to two orders of 
 magnitude from the boundaries, as observed in our universe.  With 
 an accuracy that depends on the strength of the force, multiverse 
 predictions result for both $\rho_\Lambda$ and $\bar{\rho}$ in 
 terms of the fundamental parameters that determine the location 
 of the boundaries.}
\label{fig:cc-1}
\end{center}
\end{figure}
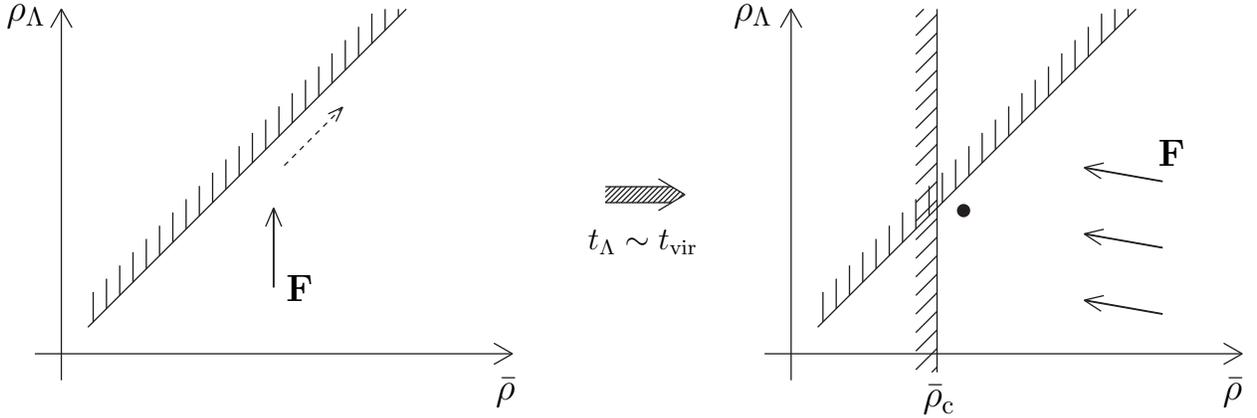

Up to logarithmic corrections from $f(M_-)$, this catastrophic boundary 
corresponds to a line of slope~$1$ in the $(\bar{\rho},\rho_\Lambda)$ 
plane shown in Fig.~\ref{fig:cc-1}.
Of course, the analysis in the following subsection would apply to any 
measure in which this line is the relevant catastrophic boundary.

\subsection{Predictions}
\label{subsec:cc1pred}

We have already assumed a force towards large vacuum energy 
$\rho_\Lambda$, and we have established two catastrophic boundaries, 
Eqs.~(\ref{eq:rho_c},~\ref{eq:rlm}).  These boundaries intersect 
at the critical point
\begin{equation}
  (\bar{\rho},\rho_\Lambda) 
  = (\bar{\rho}_{\rm c},\rho_{\Lambda,{\rm c}}),
\label{eq:2d-crit}
\end{equation}
where
\begin{equation}
  \rho_{\Lambda,{\rm c}} 
  \equiv \rho_{\Lambda,{\rm max}}(\bar{\rho}_{\rm c}) 
  = \Biggl( \frac{3^2 \pi\, c_{\rm ion} f_{\rm rad}^2}{2^6 f_\rho K^2 f_B^2} 
    \Biggr) \frac{G_{\rm N} m_e^4 m_p^2}{\alpha^4} 
  \simeq (2.8 \times 10^{-3}~{\rm eV})^4.
\label{eq:rho_Lambda-c}
\end{equation}
Here, we have used the expression for $\rho_{\rm vir,c}$ given in 
Eq.~(\ref{eq:rho_vir_c}).

Whether these critical values become predictions depends on the sign 
and magnitude of the force component in the $\bar{\rho}$ direction, 
$p_{\bar{\rho}} \equiv \partial \ln f/\partial \ln \bar{\rho}$. 
(Note that this is not the same as the force on $\bar{\rho}$ considered 
in the last section, which was understood to result from integrating 
out all other scanning parameters, such as $\Lambda$.)

The left panel of Fig.~\ref{fig:cc-1} shows the runaway problem that 
would arise in the absence of a sufficiently strong force toward small 
$\bar{\rho}$, i.e., if $p_{\bar{\rho}} > -p_{\Lambda} $.  The total force 
points to the structure formation boundary, but once this boundary is 
reached, the component tangential to the boundary goes in the positive 
$\rho_\Lambda$ and $\bar{\rho}$ directions, giving the runaway behavior 
indicated by the dashed line.  The cosmological constant problem is 
not solved in this case.  Moreover, the prediction for $\bar{\rho}$ in 
the previous section would also be lost, since the effective force on 
$\bar{\rho}$, after integrating out $\Lambda$, would point towards large 
values.

Let us assume, therefore, that $p_{\bar{\rho}} < -p_{\Lambda} $, i.e., 
the force towards small $\bar{\rho}$ is strong enough to overwhelm 
the force towards large $\rho_\Lambda$ once the catastrophic structure 
formation boundary is reached.  In this case, as shown in the right 
panel of Fig.~\ref{fig:cc-1}, the force selects the tip of the cone 
formed by the two boundaries as the most likely universe, and we 
obtain two predictions:
\begin{equation}
\begin{array}{rcccl}
  \bar{\rho} & \sim & \bar{\rho}_{\rm c} 
    & \simeq & (0.97\times 10^{-4}~{\rm eV})^4,
\label{eq-2pred} \\[2ex]
  \rho_\Lambda & \sim & \rho_{\Lambda,{\rm c}} 
    & \simeq & (2.8 \times 10^{-3}~{\rm eV})^4.
\end{array}
\end{equation}

Recall that $\bar{\rho}_{\rm c}$ and $\rho_{\rm vir,c}$, given in 
Eqs.~(\ref{eq:rho_c}) and (\ref{eq:rho_vir_c}), have the same parametric 
dependence on fundamental parameters and differ only through numerical 
factors.  Neglecting such factors, we are able to relate both 
$\bar{\rho}$ and $\rho_\Lambda$ to fundamental parameters as
\begin{equation}
  \bar{\rho} \sim \rho_\Lambda \sim \frac{G_{\rm N} m_e^4 m_p^2}{\alpha^4}.
\label{eq:barrho-Lambda-pred}
\end{equation}
As a corollary, we are able to explain the double coincidence
\begin{equation}
  t_\Lambda \sim t_{\rm rad} \sim t_{\rm vir} 
  \sim \frac{\alpha^2}{G_{\rm N} m_e^2 m_p}.
\label{eq:tcool=tvir=tlambda}
\end{equation}

Thus, we predict that most observers will find themselves in a universe 
with $\bar{\rho}$ larger, but not much larger, than $\bar{\rho}_{\rm c}$, 
and with $\rho_\Lambda$ comparable to $\bar{\rho}_{\rm vir,c}/54$. 
(Note that $\rho_\Lambda$ could be smaller or larger than this critical 
value, because the catastrophic boundary is not a line of constant 
$\rho_\Lambda$.  However, it should not differ very much from the 
critical value.)

The first prediction in Eq.~(\ref{eq-2pred}) reproduces our successful 
result in section~\ref{subsec:basic_pred}, and it yields the same 
corollaries: predictions for the characteristic mass, virial density, 
and cooling timescale of galactic halos.  The second prediction 
addresses the cosmological constant problem.  This prediction, 
too, is successful: the observed vacuum energy, $\rho_{\Lambda,o} 
= (2.3 \times 10^{-3}~{\rm eV})^4$, is only a factor $2.1$ smaller 
than the critical value.

This might suggest that the force on the cosmological constant must 
be strong.  But this is not the case.  To gauge the multiverse forces, 
we must compute the distances orthogonal to the catastrophic boundaries. 
We shall now see that this constrains the force on the cosmological 
constant to be quite weak.

\subsection{The strength of the multiverse force}
\label{subsec:cc1force}

We now have a two-dimensional parameter space, with variables
\begin{eqnarray}
  z_1 &=& \ln\frac{\bar{\rho}}{\bar{\rho}_{\rm c}},
\label{eq:z1}\\
  z_2 &=& \ln\frac{\rho_\Lambda}{\rho_{\Lambda,{\rm c}}}.
\label{eq:z2}
\end{eqnarray}
By Eq.~(\ref{eq:UBZ}), the matrix
\begin{equation}
  B = 
  \left( \begin{array}{cc}
    1 & 0 \\
    1 & -1 
  \end{array} \right)
\label{eq:B-2dim}
\end{equation}
defines two new parameters
\begin{eqnarray}
  u_1 &=& \ln\frac{\bar{\rho}}{\bar{\rho}_{\rm c}},
\label{eq:u1}\\
  u_2 &=& \ln\frac{\rho_{\Lambda,{\rm max}}(\bar{\rho})}{\rho_\Lambda},
\label{eq:u2}
\end{eqnarray}
that are orthogonal to catastrophic boundaries at $u_1=0$ and $u_2=0$, 
where $\rho_{\Lambda,{\rm max}}$ is given in Eq.~(\ref{eq:rlm}). 
(We have used that $\rho_{\Lambda,{\rm max}}$ is proportional to 
$\bar{\rho}$, up to a logarithmic term which we neglect.)

By Eq.~(\ref{eq:SBP}), $(B^{-1})^T$ defines associated force components 
$S$, with
\begin{eqnarray}
  s_1 &=& p_{\bar{\rho}} + p_{\Lambda},
\label{eq:s1-2dim}\\
  s_2 &=& -p_{\Lambda},
\label{eq:s2-2dim}
\end{eqnarray}
acting on the parameters $U$.  The absence of a runaway problem 
corresponds to the condition $s_i<0$, $i=1,2$, which agrees with the 
force conditions we identified in section~\ref{subsec:cc1pred} above.

The observed values in our universe are
\begin{eqnarray}
  u_{1,o} &\simeq& \ln 40~\, \simeq 3.7,
\label{eq:u1o-2dim}\\
  u_{2,o} &\simeq& \ln 139   \simeq 4.9.
\label{eq:u2o-2dim}
\end{eqnarray}
The second line is the statement that the cosmological constant in our 
universe is more than a hundred times smaller than necessary for large 
scale structure formation.  By demanding that these observed values 
are within one or two standard deviations from the median, we obtain 
from Eq.~(\ref{eq:snu}) that
\begin{eqnarray}
  p_{\bar{\rho}} + p_{\Lambda} &=& -\left[ 
    0.19~^{+0.31}_{-0.14}(1\sigma)~^{+0.84}_{-0.18}(2\sigma) \right],
\label{eq:2var1}\\
  -p_{\Lambda} &=& -\left[ 
    0.141~^{+0.233}_{-0.106}(1\sigma)~^{+0.626}_{-0.136}(2\sigma) \right].
\label{eq:2var2}
\end{eqnarray}

Note that the right hand side of Eq.~(\ref{eq:2var1}) is the same 
as the constraint on the force $p_{\bar{\rho}}$ derived in the 
one-dimensional setting of Eq.~(\ref{eq:force_rho-bar}).  This is 
as is should be.  The left hand side looks different, but this is 
a notational problem: in the previous section, $p_{\bar{\rho}}$ was 
understood to include all contributions from integrating out other 
parameters and boundaries.  Here, however, the contribution from 
$p_\Lambda$ was excluded in the definition of $p_{\bar{\rho}}$. 
It thus appears explicitly in Eq.~(\ref{eq:2var1}).

How do these bounds compare to theoretical expectations?  Because $0$ 
is not a special point in the spectrum of the cosmological constant, 
one can treat the statistical distribution of $\rho_\Lambda$ as flat 
in a small neighborhood of $0$ (and only an extremely small neighborhood 
is of interest anthropically):
\begin{equation}
  \frac{d \tilde{f}}{d \rho_\Lambda} = {\rm const},
\label{eq:CC-dist}
\end{equation}
at least after smoothing over intervals much smaller than the observed 
value.  Thus, the simplest theoretical expectation is a force of unit 
strength towards large values of $\Lambda$:
\begin{equation}
  \tilde{p}_\Lambda \equiv 
    \frac{\partial \ln\tilde{f}}{\partial \ln\rho_\Lambda} = 1.
\label{eq:CC-p}
\end{equation}
This result, however, is modified (in different ways) by all of the 
measures we consider here.

The catastrophic boundary we have used---the Weinberg bound on the 
formation of galactic halos---arises from the scale factor measure as 
formulated by De~Simone {\em et al.}~\cite{DeSimone:2008bq,Bousso:2008hz}. 
Correspondingly, we should here consider the implications of this 
particular measure for the strength of $F_\Lambda$.  We will compare 
this theoretical expectation to our empirical constraint, 
Eq.~(\ref{eq:2var2}).

Let $\hat{n}_{\rm obs}$ be the number density of observers per physical 
volume at the time $t_{\rm vir} + t_{\rm obs}$ when observations are 
made in a given bubble universe.  As shown in Ref.~\cite{Bousso:2008hz}, 
the scale factor measure weights observers by the physical number 
density these observers would have had at the time when the first 
structures formed that would eventually be incorporated into their 
halo.  This weight is approximately
\begin{equation}
  w_{\rm SF1} \propto 
    \left( \frac{\hat{n}_{\rm obs}}{\rho_{\rm obs}} \right) \rho_{\rm vir},
\label{eq:wsf1}
\end{equation}
where $\rho_{\rm obs}$ is the matter density at the time $t_{\rm vir} 
+ t_{\rm obs}$.  The first factor is the number of observers per unit 
mass.  In our sharp-boundary approximation, this factor is constant 
and nonzero on the anthropic side of the catastrophic boundaries in 
the $(\bar{\rho},\rho_\Lambda)$ plane.  Beyond the cooling and structure 
boundaries, it vanishes.

The second factor, up to numerical coefficients, agrees with 
$\bar{\rho}$, so the weight $w_{\rm SF1}$ contributes a single 
power of $\bar{\rho}$ to the distribution function $f$.  Thus, 
the force on the anthropic side receives a contribution
\begin{equation}
  (\Delta p_{\bar{\rho}}, \Delta p_\Lambda)_{\rm SF1} = (1,0)
\label{eq:SF1-p}
\end{equation}
from the measure.  Therefore, the total force is $(p_{\bar{\rho}}, 
p_{\Lambda}) = (1+\tilde{p}_{\bar{\rho}}, 1)$, where 
$\tilde{p}_{\bar{\rho}} \equiv \partial \ln\tilde{f}/\partial 
\ln\bar{\rho}$ is the unknown contribution from the distribution 
of $\bar{\rho}$ among landscape vacua.  For the linear combinations 
constrained in Eqs.~(\ref{eq:2var1},~\ref{eq:2var2}), we find
\begin{equation}
  (p_{\bar{\rho}}+p_\Lambda, p_{\Lambda}) = (2+\tilde{p}_{\bar{\rho}}, 1).
\label{eq:SF1-p_2}
\end{equation}

The first constraint can be satisfied by assuming that the unknown 
prior landscape force, $\tilde{p}_{\bar{\rho}}$, lies between $-2$ 
and $-3$.  While this may seem like an implausibly strong force 
towards small virial density---an inverse power law of mass dimension 
$-8$ to $-12$---we cannot rule out that it may be realized in the 
string landscape.

The second constraint, however, on the force on $\Lambda$, is 
definitely incompatible with the theoretical expectation at the 
$2\sigma$ level. (It would be compatible at the $3\sigma$ level.) 
Thus, with the scale factor measure in the formulation of De~Simone 
{\em et al.}~\cite{DeSimone:2008bq}, multiverse forces cannot 
simultaneously explain $\rho_{\rm vir}$ and $\rho_\Lambda$ without 
additional assumptions.

\section{Predicting the Cosmological Constant and Observer Time Scale}
\label{sec:CC-2}

In this section, we will again consider the case of scanning both the 
cosmological constant $\rho_\Lambda$ and the virial density parameter 
$\bar{\rho}$, but using a different measure for regulating the 
divergences of the eternally inflating multiverse.  In the previous 
section we used the scale factor measure in the formulation 
of Ref.~\cite{DeSimone:2008bq} (SF1).  We will now explore the 
implications of the causal patch (CP)~\cite{Bousso:2006ev}, causal 
diamond (CD)~\cite{Bousso:2007kq}, and modified scale factor 
(SF2)~\cite{Bousso:2008hz} measures.  We will find important 
differences to the previous section, and subtle differences among 
the three measures we now consider:
\begin{itemize}
\item All three measures yield the same set of three catastrophic 
boundaries, in the three-dimensional parameter space $(\bar\rho, 
\rho_\Lambda , t_{\rm obs})$.  The first two boundaries are the 
same as for the SF1 measure; the boundary on $t_{\rm obs}$ is new.
\item Because of the appearance of $t_{\rm obs}$, we will 
allow three parameters to vary in this section: $\bar{\rho}$, 
$\rho_\Lambda$, and $t_{\rm obs}$.
\item The three measures make different contributions to the 
multiverse force vector.  We will find observational constraints on 
the corresponding multiverse forces that are in excellent (CP, CD) 
or reasonable (SF2) agreement with theoretical expectations.
\end{itemize}

\subsection{Catastrophic boundary from galaxy dispersion}
\label{subsec:cat-gal-dis}

All three measures considered in this section exponentially suppress 
the weight of observations that take place after the cosmological 
constant dominates, like $\exp{(-3t/t_\Lambda)}$.  There is some 
ambiguity about which critical value to pick in the sharp-boundary 
approximation.  We will use $t_\Lambda \equiv (8\pi G_{\rm N} 
\rho_\Lambda/3)^{-1/2}$, but our results do not depend sensitively 
on small modifications of this choice.

Thus, in the sharp boundary approximation, we require
\begin{equation}
  t_{\rm vir} + t_{\rm obs} < t_\Lambda,
\label{eq:ttt}
\end{equation}
as a necessary condition for the existence of observers.  Here, we 
define $t_{\rm obs}$ as the time delay between the formation of a 
galactic halo at time $t_{\rm vir}$ and the time when the halo hosts 
observers.  (We neglect the cooling time scale, which must not exceed 
$t_{\rm vir}$ in any case.)  The delay $t_{\rm obs}$ includes the 
time needed for the formation of a solar system and Darwinian evolution, 
or whatever other forms the development of life in gravitationally 
collapsed structure may take.  We have no idea how to relate 
$t_{\rm obs} $ to fundamental particle physics or cosmological 
parameters, but we will assume that it depends on sufficiently 
many parameters, sufficiently sensitively, that we can treat it 
as scanning independently of the other four parameters we are varying 
in this paper ($\bar{\rho}$, $t_{\rm rad}$, $\rho_\Lambda$, and $Q$).%
\footnote{Note that our treatment of $t_{\rm obs}$ is not very 
 different from $t_\Lambda$.  Neither quantity can be simply computed 
 from ``fundamental'' parameters of our vacuum, but both are expected 
 to scan in the multiverse.  We predict both quantities by correlating 
 them with other scales.}

In the presence of power law forces on $\bar{\rho}$ and $t_{\rm obs}$, 
we make a negligible error if we write Eq.~(\ref{eq:ttt}) as two 
independent catastrophic boundaries:
\begin{eqnarray}
  t_{\rm obs} &<& t_\Lambda,
\label{eq:totl}\\
  t_{\rm vir} &<& t_\Lambda.
\label{eq:tvtl0}
\end{eqnarray}
The existence of galaxies, of course, remains a necessary condition 
for observers, and it yields a bound tighter than Eq.~(\ref{eq:tvtl0}): 
from Eq.~(\ref{eq:rlm}) we find
\begin{equation}
  t_{\rm vir} < \frac{\pi}{3^{3/2}} t_\Lambda \simeq 0.60\, t_\Lambda.
\label{eq:tvtl}
\end{equation}
Thus, we will use Eq.~(\ref{eq:totl}) as the new catastrophic boundary 
for the new parameter $t_{\rm obs}$.  We will continue to treat 
$\rho_{\rm vir}/54$, from Eq.~(\ref{eq:rlm}), as a catastrophic 
upper bound on $\rho_\Lambda$, and $\bar{\rho}_{\rm c}$, given in 
Eq.~(\ref{eq:rho_c}), as a catastrophic lower bound on $\bar{\rho}$.

\subsection{Predictions}
\label{subsec:3dpredictions}

The three boundaries we have established intersect at the critical point
\begin{equation}
  (\bar{\rho}, \rho_\Lambda , t_{\rm obs}) 
  = (\bar{\rho}_{\rm c}, \rho_{\Lambda,{\rm c}}, t_{\rm obs,c}),
\label{eq:3dim-crit}
\end{equation}
where
\begin{equation}
  t_{\rm obs,c} = t_\Lambda(\rho_{\Lambda,{\rm c}}) 
  = \frac{2^{3/2} f_\rho^{1/2} K f_B}{3^{1/2}\pi\, c_{\rm ion}^{1/2} 
    f_{\rm rad}}\, \frac{\alpha^2}{G_{\rm N} m_e^2 m_p}\simeq 11.0~{\rm Gyr}.
\label{eq:t_obs-c}
\end{equation}

Supposing that the three-dimensional multiverse force points towards 
this critical point, we predict that observed values of these parameters 
should be within one or two orders of magnitude of the critical values. 
The predictions for $\bar{\rho}$ and $\rho_\Lambda $ are the same as 
in the previous sections.  The new prediction for $t_{\rm obs}$ is 
that it should not differ much from $t_{\rm obs,c}$, given above:
\begin{equation}
  t_{\rm obs} \sim t_{\rm obs,c}.
\label{eq:t_obs-pred}
\end{equation}
The observed value is
\begin{equation}
  t_{\rm obs,o} = 10.1~{\rm Gyr},
\label{eq:t_obs-o}
\end{equation}
so this third basic prediction is also successful.

We have now explained the triple coincidence
\begin{equation}
  t_{\rm obs} \sim t_\Lambda \sim t_{\rm rad} 
  \sim t_{\rm vir} \sim \frac{\alpha^2}{G_{\rm N} m_e^2 m_p}.
\label{eq:triple}
\end{equation}
Next, we will determine under which conditions the multiverse force 
points towards the critical point, and we will establish empirical 
constraints on the the three force components.

\subsection{Forces with the causal patch measure}
\label{subsec:3dforces}

In addition to the variables $z_1$ and $z_2$ defined in 
section~\ref{subsec:cc1force}, we now have the third scanning 
parameter
\begin{equation}
  z_3 = \ln \frac{t_{\rm obs}}{t_{\rm obs,c}}.
\label{eq:z3}
\end{equation}
The matrix
\begin{equation}
  B = 
  \left( \begin{array}{ccc}
    1 &  0 & 0 \\
    1 & -1 & 0 \\
    0 & -1 & -2 
  \end{array} \right)
\label{eq:B-3dim}
\end{equation}
defines the boundary-orthogonal parameter
\begin{equation}
  u_3 = \ln \left( \frac{3}{8\pi G_{\rm N}}\, 
    \frac{1}{\rho_\Lambda t_{\rm obs}^2} \right).
\label{eq:u3}
\end{equation}

By Eq.~(\ref{eq:SBP}), $(B^{-1})^T$ defines a boundary-orthogonal 
force vector $S$, with
\begin{eqnarray}
  s_1 &=& p_{\bar{\rho}} + p_{\Lambda} - \frac{1}{2}\, p_{\rm obs},
\label{eq:s1-3dim}\\
  s_2 &=& -p_{\Lambda} + \frac{1}{2}\, p_{\rm obs},
\label{eq:s2-3dim}\\
  s_3 &=& -\frac{1}{2}\, p_{\rm obs},
\label{eq:s3-3dim}
\end{eqnarray}
where $p_{\rm obs} = \partial \ln f/\partial \ln t_{\rm obs}$. 
The absence of a runaway problem, i.e., the condition that the 
multiverse force points towards the critical point, is that all 
$s_i < 0$.  In particular, we learn that it is necessary to assume 
a statistical preference for a large time scale for the development 
of observers in gravitationally bound regions.  This assumption 
seems plausible, given the complex processes that are likely to 
be involved.  Since $p_{\Lambda}$ appears only in the combination 
$p_\Lambda - \frac{1}{2} p_{\rm obs}$, this preference acts counter 
to the multiverse force towards large cosmological constant.  This, 
too, is intuitively plausible: there are fewer vacua with smaller 
cosmological constant, but on average they contain more observers.

From the observed values in our universe we find
\begin{eqnarray}
  u_{1,o} &\simeq& \ln 40~\,\simeq 3.7,
\label{eq:u1o-3dim}\\
  u_{2,o} &\simeq& \ln 139  \simeq 4.9,
\label{eq:u2o-3dim}\\
  u_{3,o} &\simeq& \ln 2.5  \simeq 0.92.
\label{eq:u3o-3dim}
\end{eqnarray}
The fact that $u_3$ is of order unity is closely related to the 
coincidence problem: we live in the unique era when vacuum and matter 
energy densities are comparable.  By demanding that these observed 
values are within one or two standard deviations from the median, 
we obtain from Eq.~(\ref{eq:snu}) that
\begin{eqnarray}
  p_{\bar{\rho}} + p_{\Lambda} - \frac{1}{2}\, p_{\rm obs} &=& -\left[ 
    0.19~^{+0.31}_{-0.14}(1\sigma)~^{+0.84}_{-0.18}(2\sigma) \right],
\label{eq:3var1}\\
  -p_{\Lambda} + \frac{1}{2}\, p_{\rm obs} &=& -\left[ 
    0.141~^{+0.233}_{-0.106}(1\sigma)~^{+0.626}_{-0.136}(2\sigma) \right],
\label{eq:3var2}\\
  -\frac{1}{2}\, p_{\rm obs} &=& -\left[ 
    0.76~^{+1.25}_{-0.57}(1\sigma)~^{+3.37}_{-0.73}(2\sigma)\right].
\label{eq:3var3}
\end{eqnarray}
The first two constraints are identical to those found in the previous 
section, except that the contribution to the multiverse force from 
$p_{\rm obs}$ is now made explicit on the left hand side.

How do these bounds compare to theoretical expectations?  As reviewed 
in the previous section, the statistical distribution of $\Lambda$ in 
the landscape is flat, giving a prior force $\tilde{p}_\Lambda = 1$ 
towards large values.  As discussed at the beginning of this section, 
the three measures differ in the spatial size of the cutoff region. 
On the anthropic side of the catastrophic boundary, the number 
of observers included depends differently on $t_{\rm obs}$ and 
$\rho_\Lambda$ for each measure.  This means that the measures make 
different contributions to the multiverse force vector.  We will 
discuss each measure in turn, beginning with the CP measure.

The CP measure contributes a factor 
$\rho_\Lambda^{-1/2}$~\cite{Bousso:2007kq}:
\begin{equation}
  w_{\rm CP} = \left(\frac{\hat{n}_{\rm obs}}{\rho_{\rm obs}}\right) 
    \rho_\Lambda^{-1/2},
\label{eq:wcp}
\end{equation}
where the first factor, the density of observers per unit mass, is 
again approximated as constant on the anthropic side of the catastrophic 
boundaries and zero beyond.  Thus, the CP measure makes no contribution 
to the forces on $\bar{\rho}$ and $t_{\rm obs}$, and we find:
\begin{equation}
  (\Delta p_{\bar{\rho}}, \Delta p_\Lambda, \Delta p_{\rm obs})_{\rm CP} 
  = (0, -\frac{1}{2}, 0).
\label{eq:CP-p}
\end{equation}

The CP measure, therefore, leads to a reduced force on the cosmological 
constant:
\begin{equation}
  p_\Lambda = \frac{1}{2}.
\label{eq:plcp}
\end{equation}
In Eq.~(\ref{eq:3var2}), $p_\Lambda$ and the force towards large 
observer time scale appear with opposite sign, reducing the magnitude 
of the left hand side further.  This should be compared to the analogous 
equation Eq.~(\ref{eq:2var2}) for the SF1 measure, which contained 
only $p_\Lambda = 1$, and so could not be satisfied at the $2\sigma$ 
level.  By contrast, the above three empirical constraints are all 
compatible with Eq.~(\ref{eq:plcp}) at the $1\sigma$ level.  Possible 
values for $(p_{\bar{\rho}}, p_{\rm obs})$ include $(-\frac{1}{2}, 
\frac{1}{2})$ and $(-\frac{1}{4}, \frac{3}{4})$.

\subsection{Forces with the causal diamond measure}
\label{subsec:force-mod-CD}

How would this discussion be modified if we used the CD measure? 
The catastrophic boundaries remain unchanged, preserving both the 
general prediction that we should find ourselves near the triple 
critical point $(\bar{\rho}_{\rm c}, \rho_{\Lambda,{\rm c}}, 
t_{\rm obs,c})$, and the empirical constraints given in 
Eqs.~(\ref{eq:3var1}~--~\ref{eq:3var3}).

However, the measure contributes differently to the multiverse force. 
It does not contribute to the force on $\rho_\Lambda$, but instead 
favors late times of observation, $t_{\rm vir} + t_{\rm obs}$, because 
the mass inside the causal diamond depends linearly on this quantity 
as long as it does not exceed $t_\Lambda $.  This corresponds to a 
contribution of a force of strength $1$ on the variable $t_{\rm vir} 
+ t_{\rm obs}$:
\begin{equation}
  \Delta p_{\rm vir+obs} = 1.
\label{eq:eq:pvocd}
\end{equation}
In terms of the individual variables, $t_{\rm vir}$ and $t_{\rm obs}$, 
this implies that for $t_{\rm vir} \gg t_{\rm obs}$, the CD 
measure contributes $(\Delta p_{\bar{\rho}}, \Delta p_\Lambda, 
\Delta p_{\rm obs})_{\rm CD} = (-\frac{1}{2}, 0, 0)$; in the 
opposite limit, $(0,0,1)$.

Thus, the force is not uniform in the allowed quadrant of parameter 
space; instead, it changes across the hypersurface $t_{\rm vir} \sim 
t_{\rm obs}$.  This case cannot be treated with the simple analysis 
following Eq.~(\ref{eq:fz}), and we will not attempt here to constrain 
the prior force strengths and directions compatible with the observed 
values of the three parameters.

We note, instead, that the measure force $\Delta p_{\rm vir+obs}$ acts 
to increase $t_{\rm vir} + t_{\rm obs}$.  Because of the constraint 
Eq.~(\ref{eq:ttt}) that arises in this measure, it also acts to increase 
$t_\Lambda$ with strength $1$, or equivalently to decrease $\Lambda$ 
with strength $-1/2$.  Therefore, the measure force effectively 
counteracts the prior pressure towards large $\Lambda$, reducing it 
from $1$ to $1/2$.  This is similar to the causal patch, where this 
reduction happens more directly, without the need to invoke one of 
the catastrophic boundaries; see Eqs.~(\ref{eq:CP-p},~\ref{eq:plcp}). 
We thus expect that the causal diamond, like the patch, leads to 
a good agreement between observed and predicted values for these 
three parameters, for a reasonable range of prior force strengths 
$(p_{\bar{\rho}}, p_{\rm obs})$.

\subsection{Forces with the modified scale factor measure}
\label{subsec:force-mod-SF}

Finally, let us turn to SF2, a version of the scale factor measure that 
requires $t_{\rm vir} + t_{\rm obs} < t_\Lambda$~\cite{Bousso:2008hz}. 
We may again translate this condition into the two boundaries given in
Eqs.~(\ref{eq:rlm},~\ref{eq:totl}).%
\footnote{Strictly, the SF2 measure involves an effective $t_{\rm obs}$ 
 related to the time when the latest geodesic reaching observers 
 achieved maximum expansion.  One would expect that this occurs 
 at about one half of the time $t_{\rm vir} + t_{\rm obs}$, so one 
 should redefine $t_{\rm obs,SF2} = (t_{\rm obs,CP} - t_{\rm vir})/2$. 
 However, compared to the CP, the SF2 measure suppresses large values 
 of $\rho_\Lambda \gtrsim t_{\rm obs}^{-2}/G_N$ by an additional factor 
 $\rho_\Lambda^{-3/2}$; so the sharp boundary on $t_{\rm obs}$, which 
 was $t_{\rm obs,CP} < t_\Lambda$, should also be moved to a smaller 
 value.  If we choose $t_{\rm obs,SF2} < t_\Lambda/2$, then these 
 two effects roughly cancel, so we will omit both from our analysis 
 for simplicity.}
Thus, the catastrophic boundaries are the same as for the other 
two measures considered in this section.  And so the SF2 measure, 
too, gives rise to the general prediction that we should find 
ourselves near the triple critical point $(\bar{\rho}_{\rm c}, 
\rho_{\Lambda,{\rm c}}, t_{\rm obs,c})$, and to the empirical 
constraints given in Eqs.~(\ref{eq:3var1}~--~\ref{eq:3var3}).

However, the SF2 measure makes contributions to the forces that 
differ strongly from those arising from the CP or CD.  It weights 
by the physical density of observers, which is proportional to 
$(t_{\rm vir} + t_{\rm obs})^{-2}$.  Thus, for $t_{\rm vir} \gg 
t_{\rm obs}$, it contributes $(\Delta p_{\bar{\rho}}, \Delta p_\Lambda, 
\Delta p_{\rm obs})_{\rm SF2} = (1, 0, 0)$; in the opposite limit, 
$(0,0,-2)$.  This is unhelpful, because both contributions go 
in the wrong direction: by Eq.~(\ref{eq:3var3}), $p_{\rm obs}$ 
must be positive, and by Eqs.~(\ref{eq:3var1},~\ref{eq:3var2}), 
$p_{\bar{\rho}}$ must be negative.  Thus, the empirical constraints 
require that large landscape forces $\tilde{p}_{\bar{\rho}} < 0$ 
and $\tilde{p}_{\rm obs} > 0$ overcompensate for the contribution 
from the measure.  This is particularly acute for $p_{\rm obs}$, 
which must be larger than in the CP measure by about $1$, in order 
to compensate for the larger value of $p_\Lambda$ ($1$ instead 
of $1/2$) in Eq.~(\ref{eq:3var2}).  But we cannot exclude the 
SF2 measure; it is possible that the prior distributions in the 
landscape satisfy these constraints.

\section{Predicting the Primordial Density Contrast}
\label{sec:comp}

Our halo, which is among the largest galactic halos, virialized 
approximately $3~{\rm Gyr}$ after the big bang.  According to 
Eq.~(\ref{eq:tcompmax}), Compton cooling in our universe became 
ineffective only a little earlier, at $360$ million years.  We 
have argued that there is an environmental understanding of why 
the Milky Way halo was just able to cool radiatively, could it be 
that there is also an environmental reason to explain why it just 
failed to cool by inverse Compton scattering?

Indeed, the proximity to Compton cooling is striking.  For example, 
with $Q$ a factor of $4$ larger than in our universe (and holding fixed 
the temperature at equality), a halo with the mass of the Milky Way 
would have undergone Compton cooling.  This suggests that there is 
a catastrophic boundary associated with Compton cooling, and that the 
multiverse distribution has pushed us close to it.  In this section, 
we will explore the consequences of such a boundary.  We stress that 
we do not explain why the anthropic weighting factor varies suddenly 
across the Compton boundary; we will simply pursue the consequences 
of {\em assuming} that it does.

\subsection{Catastrophic boundary from Compton cooling}
\label{subsec:cat-Comp}

As discussed in section~\ref{subsec:compton}, Compton cooling is 
effective if $t_{\rm comp} < t_{\rm vir}$; this is the case for all 
galaxies that form prior to the cosmological time $t_{\rm comp,max}$ 
given in Eq.~(\ref{eq:tcompmax}).  We are assuming that Compton 
cooling inhibits the formation of galaxies, or changes their structure 
sufficiently to make them inhabitable.  In our universe, Compton 
cooling only affects halos that virialize very early and thus have 
small mass.  We have defined a catastrophe as the absence of habitable 
galaxies of any mass scale.  Thus, a catastrophe occurs if parameters 
are altered so that Compton cooling affects {\em all\/} halos, 
including those of the largest mass able to cool radiatively, $M_+$, 
given in Eq.~(\ref{eq:M_max}).  Namely, a catastrophe is averted if
\begin{equation}
  t_{\rm comp,max} < t_{\rm vir} (M_+).
\label{eq:tctv}
\end{equation}
Using Eq.~(\ref{eq:t_vir}), we thus find the catastrophic boundary
\begin{equation}
  Q^2 \bar{\rho} < \frac{3^6 \times 5\, N_{\rm eq}\, c_{\rm vir}^{10/3} 
    \delta_{\rm col}^{5}}{2^{10}\, \pi^{1/3} f(M_+)^{5}}\, 
    \frac{G_{\rm N} m_e^6}{\alpha^4}.
\label{eq:qrho}
\end{equation}
We could have expressed this boundary in terms of any two of the three 
parameters $Q$, $T_{\rm eq}$, and $\bar{\rho} = Q^3 T_{\rm eq}^4$. 
For continuity with the previous sections, we have chosen to retain 
$\bar{\rho}$ as a scanning parameter.

There is a subtlety here.  Compton cooling can be catastrophic only 
if it is faster than radiative cooling; otherwise, it would have no 
opportunity to act.  However, this condition will automatically be 
satisfied if the above inequality is violated.  For the largest halos 
able to cool radiatively, $t_{\rm rad} = t_{\rm vir}$ by definition, 
so a violation of Eq.~(\ref{eq:tctv}) implies that Compton cooling 
is faster than radiative cooling.  For smaller halos, which virialize 
earlier, $t_{\rm comp} \propto t_{\rm vir}^{8/3}$ whereas $t_{\rm rad} 
\propto t_{\rm vir}^{5/3}$.  Therefore, Compton cooling will be 
faster than radiative cooling for all halos capable of cooling 
if Eq.~(\ref{eq:tctv}) is violated.

\subsection{Predictions}
\label{subsec:Comp-pred}

We can now consider the four-dimensional parameter space $(\bar{\rho}, 
\rho_\Lambda, t_{\rm obs}, Q)$.  We have established four catastrophic 
boundaries, which intersect at the critical point
\begin{equation}
  (\bar{\rho}_{\rm c}, \rho_{\Lambda,{\rm c}}, t_{\rm obs,c}, Q_{\rm c}).
\label{eq:4d-crit}
\end{equation}
The first three critical values have been derived in the three previous 
sections; the fourth can be obtained by setting $\bar{\rho} \to 
\bar{\rho}_{\rm c}$ in Eq.~(\ref{eq:qrho}).  Using Eq.~(\ref{eq:rho_c}), 
we find
\begin{equation}
  Q_{\rm c} = \frac{\pi^{1/3} c_{\rm vir} \delta_{\rm col} N_{\rm eq} 
    f_\rho K f_B}{8 f(M_{\rm c})\, c_{\rm ion}^{1/2}\, f_{\rm rad}}\, 
    \frac{m_e}{m_p} 
  \simeq 2.4 \times 10^{-3}.
\label{eq:Qc}
\end{equation}

As usual, let us now assume that the four-dimensional multiverse force 
is directed towards the critical point.  (We will analyze this condition 
in detail in the next subsection.)  Then we predict
\begin{equation}
  Q \sim Q_{\rm c},
\label{eq:Q-pred}
\end{equation}
in addition to the three other predictions $(\bar{\rho}, \rho_{\Lambda}, 
t_{\rm obs}) \sim (\bar{\rho}_{\rm c}, \rho_{\Lambda,{\rm c}}, 
t_{\rm obs,c})$ established in the previous sections.  The observed 
value is $Q_o \simeq 2.0 \times 10^{-5}$, a factor of $120$ below 
the critical value.  As a corollary, we are now able to explain the 
quadruple coincidence
\begin{equation}
  t_{\rm comp,max} \sim t_{\rm obs} \sim t_\Lambda \sim t_{\rm rad} 
  \sim t_{\rm vir} \sim \frac{\alpha^2}{G_{\rm N} m_e^2 m_p}
\label{eq:quadruple}
\end{equation}
of timescales observed in our universe.

\subsection{Forces}
\label{subsec:4dforces}

In addition to the variables $z_1, z_2, z_3$ defined in 
section~\ref{subsec:3dforces}, we now have the forth scanning 
parameter
\begin{equation}
  z_4 = \ln\frac{Q}{Q_{\rm c}}.
\label{eq:z4}
\end{equation}
The matrix
\begin{equation}
  B = 
  \left( \begin{array}{cccc}
    1  &  0 &  0 &  0 \\
    1  & -1 &  0 &  0 \\
    0  & -1 & -2 &  0 \\
    -1 &  0 &  0 & -2 
  \end{array} \right)
\label{eq:B-4dim}
\end{equation}
defines the boundary-orthogonal parameter
\begin{equation}
  u_4 = \ln \frac{Q_{\rm c}^2\, \bar{\rho}_{\rm c}}{Q^2\, \bar{\rho}},
\label{eq:u4}
\end{equation}
and leaves the definitions of $u_1, u_2, u_3$ unchanged compared 
to section~\ref{subsec:3dforces}.

By Eq.~(\ref{eq:SBP}), $(B^{-1})^T$ defines a boundary-orthogonal 
force vector $S$, with
\begin{eqnarray}
  s_1 &=& p_{\bar{\rho}} + p_{\Lambda} - \frac{1}{2}\, p_{\rm obs} 
    -\frac{1}{2}\, p_{Q},
\label{eq:s1-4dim}\\
  s_2 &=& -p_{\Lambda} + \frac{1}{2}\, p_{\rm obs},
\label{eq:s2-4dim}\\
  s_3 &=& -\frac{1}{2}\, p_{\rm obs},
\label{eq:s3-4dim}\\
  s_4 &=& -\frac{1}{2} p_Q.
\label{eq:s4-4dim}
\end{eqnarray}
The absence of a runaway problem, i.e., the condition that the 
multiverse force points towards the critical point, is that all 
$s_i < 0$.  In particular, we must assume a statistical preference 
for large values of $Q$, which seems entirely natural.

From the observed values in our universe we find
\begin{eqnarray}
  u_{1,o} &\simeq& \ln 40~\, \simeq 3.7,
\label{eq:u1o-4dim}\\
  u_{2,o} &\simeq& \ln 139   \simeq 4.9,
\label{eq:u2o-4dim}\\
  u_{3,o} &\simeq& \ln 2.5   \simeq 0.92,
\label{eq:u3o-4dim}\\
  u_{4,o} &\simeq& \ln 355   \simeq 5.9.
\label{eq:u4o-4dim}
\end{eqnarray}
By demanding that these observed values are within one or two standard 
deviations from the median, we obtain from Eq.~(\ref{eq:snu}) that
\begin{eqnarray}
  p_{\bar{\rho}} + p_{\Lambda} - \frac{1}{2}\, p_{\rm obs} 
    - \frac{1}{2}\, p_{Q} &=& -\left[ 
    0.19~^{+0.31}_{-0.14}(1\sigma)~^{+0.84}_{-0.18}(2\sigma) \right],
\label{eq:4var1}\\
  -p_{\Lambda} + \frac{1}{2}\, p_{\rm obs} &=& -\left[ 
    0.141~^{+0.233}_{-0.106}(1\sigma)~^{+0.626}_{-0.136}(2\sigma)\right],
\label{eq:4var2}\\
  -\frac{1}{2}\, p_{\rm obs} &=& -\left[ 
    0.76~^{+1.25}_{-0.57}(1\sigma)~^{+3.37}_{-0.73}(2\sigma)\right],
\label{eq:4var3}\\
  -\frac{1}{2}\, p_{Q} &=& -\left[ 
    0.12~^{+0.20}_{-0.09}(1\sigma)~^{+0.53}_{-0.11}(2\sigma)\right].
\label{eq:4var4}
\end{eqnarray}
The first three constraints are identical to those found in 
section~\ref{subsec:3dforces}, except that the contribution 
to the multiverse force from $p_Q$ is now made explicit on 
the left hand side.

How do these bounds compare to theoretical expectations?  We will 
discuss only the causal patch measure here, which leads to the force 
$p_\Lambda = \frac{1}{2}$ on the cosmological constant.  With this 
value, the above constraints can easily be satisfied at the $1\sigma$ 
level with natural force strengths, e.g., with $(p_{\bar{\rho}}, 
p_{\rm obs}, p_Q) = (-\frac{1}{2}, \frac{1}{2}, \frac{1}{2})$, 
$(-\frac{1}{4}, \frac{1}{2}, \frac{1}{4})$ or 
$(-\frac{1}{4}, \frac{3}{4}, \frac{1}{2})$.

\section{Discussion}
\label{sec:concl}

We have found that multiverse forces and catastrophic boundaries can 
explain the quadruple coincidence of timescales, Eq.~(\ref{eq:coscoinc}), 
or equivalently, the observed values of the virial density, the 
cosmological constant, the observer timescale, and the timescales 
for radiative and Compton cooling.

As more parameters are taken to scan, dare we think that all fundamental 
parameters could be determined in this way?  The answer is clearly no: 
as far as we know many parameters, such as the strong $CP$ parameter 
and the $\tau$ lepton mass, are not relevant for any catastrophic 
boundary.  Still, could it be that all the parameters that do enter 
catastrophic boundaries get determined environmentally?  For the 
boundaries considered so far, the relevant set of parameters is
\begin{equation}
  \alpha, m_e, m_p, G_{\rm N}, Q, T_{\rm eq}, \rho_\Lambda, \eta_B,
\label{eq:param}
\end{equation}
where $\eta_B$ is the number density of baryons relative to photons. 
Depending on the measure used to compute the statistical averages 
we could also include $t_{\rm obs}$.  If all the parameters of 
Eq.~(\ref{eq:param}) scan independently then the structure and 
cooling boundaries can determine $\rho_\Lambda$, $Q$, $T_{\rm eq}$ 
(and $t_{\rm obs}$), but what of the others?

The ratio $m_e/m_p$ plays an important role in the boundaries of nuclear 
stability, leaving $\alpha, m_p, \eta_B$ and $G_{\rm N}$.  The proton 
mass is directly derived from the QCD scale, $\Lambda_{\rm QCD}$, 
and we can assume that this is related to $\alpha$ by gauge coupling 
unification%
\footnote{The relation involves the weak scale, which may be determined 
 by some other boundary.}
--- i.e. we are not quite right to say that all the parameters of 
Eq.~(\ref{eq:param}) scan independently.  Taking $G_{\rm N}$ to 
set the unit of energy, $M_{\rm Pl}$, leaves just two undetermined 
dimensionless parameters: $\eta_B$ and $\Lambda_{\rm QCD}/M_{\rm Pl}$. 
Thus the landscape force and catastrophic boundaries have rigidly 
determined the entire parameter set of Eq.~(\ref{eq:param}), with 
the exception of the baryon asymmetry and an overall mass hierarchy. 
In fact, they may overdetermine it if $t_{\rm obs}$ is a function of 
the parameters listed in Eq.~(\ref{eq:param}).   This, however, is not 
the case if $t_{\rm obs}$ is very sensitive to the values of $\alpha$, 
$m_e$, $\cdots$, so that it can take a wide range of values within 
the parameter space in which the fundamental parameters, $\alpha$, 
$m_e$, $\cdots$, satisfy the nuclear stability and other requirements.
In such a case, $t_{\rm obs}$ can be viewed as an independent 
scanning parameter.

Our universe lies at the tip of a cone in the four dimensional parameter 
space of $\{m_e/m_p, Q, T_{\rm eq}, \rho_\Lambda\}$ (or of five dimensional 
space including $t_{\rm obs}$); but when an additional dimension such 
as $\eta_B$ and $\Lambda_{\rm QCD}/M_{\rm Pl}$ is added, a runaway 
behavior could reemerge.  It is, however, possible that the direction 
opened by the addition of a new parameter is again constrained by a 
new catastrophic boundary containing that parameter.  For example, the 
up and down quark masses are constrained by further nuclear stability 
boundaries, and the weak scale and baryon asymmetry play a crucial role 
at the boundary associated with big bang nucleosynthesis.  Given these 
possible boundaries, we can imagine that most of the environmentally 
relevant parameters in the standard particle physics and astrophysics 
are determined by multiverse physics as described here.

Ultimately, an important question will be what determines the overall 
mass hierarchy between the Planck and particle physics mass scales, 
allowing the existence of large observable universe.  We see three 
different possibilities:
\begin{itemize}
\item The multiverse distribution function.  For example, the unified 
coupling at the Planck scale may have a distribution $f(\alpha)$ 
peaked around some perturbative value that corresponds to a peak in the 
distribution for $\Lambda_{\rm QCD}$ exponentially below $M_{\rm Pl}$.
\item A further catastrophic boundary.  As $\Lambda_{\rm QCD}/M_{\rm Pl}$ 
grows, complex structures such as the sun will get smaller, and eventually 
cause catastrophic changes.  Perhaps our universe is already close to 
some such boundary.  In this case, $n$ catastrophic boundaries constrain 
$n$ scanning parameters. (Throughout this paper we have assumed a multiverse 
force that places typical observers in universes near the tip of a cone. 
An alternative possibility is that the observer region in the $n$ dimensional 
space is very small.)
\item The discretuum.  The vacuum energy scales as the 6th power of the 
overall hierarchy: $\rho_\Lambda \propto (\Lambda_{\rm QCD}/M_{\rm Pl})^6$, 
given that environmental selection by nuclear stability sets $m_e \sim 
m_p \sim \Lambda_{\rm QCD}$.  Suppose that $f(\alpha)$ gives a force 
to low values of $(\Lambda_{\rm QCD}/M_{\rm Pl})$.  The vacuum energy 
may become so small that the discreteness of the string landscape will 
become apparent, eventually stopping the potential runaway.  If this 
is the origin of the overall hierarchy, the number of string vacua 
is of order $M_{\rm Pl}^4/\rho_\Lambda \approx 10^{120}$.
\end{itemize}
At present we are unable to compute the overall hierarchy, no matter how 
it is determined.  Nevertheless, we are encouraged that simple arguments 
based on galaxy cooling allow a broad numerical understanding of 
$\rho_\Lambda$, $M_{\rm gal}$, $Q$ and $T_{\rm eq}$ in terms of the 
fundamental parameters in particle physics, $\alpha$, $m_e$, $m_p$ 
and $G_{\rm N}$.

\section*{Acknowledgments}

We thank Roni Harnik, Stefan Leichenauer and Jens Niemeyer for discussions. 
This work was supported in part by the Director, Office of Science, 
Office of High Energy and Nuclear Physics, of the US Department of Energy 
under Contract DE-AC02-05CH11231, and in part by the National Science 
Foundation under grant PHY-0457315.  The work of R.B. was supported 
in part by a Career grant of the National Science Foundation.  The work 
of Y.N. was supported in part by the National Science Foundation under 
grant PHY-0555661 and by the Alfred P. Sloan Foundation.



\begin{thebibliography}{0}

\bibitem{Dyson:1979zz}
F.~J.~Dyson,
Rev.\ Mod.\ Phys.\  {\bf 51}, 447 (1979).

\bibitem{Weinberg:1987dv}
S.~Weinberg,
Phys.\ Rev.\ Lett.\  {\bf 59}, 2607 (1987).

\bibitem{Barrow-Tipler}
J.~D.~Barrow and F.~J.~Tipler,
{\it The Anthropic Cosmological Principle}
(Oxford University Press, Oxford, United Kingdom, 1986).

\bibitem{Efstathiou:1995ne}
G.~Efstathiou,
Mon.\ Not.\ Roy.\ Astron.\ Soc.\  {\bf 274}, L73 (1995);
H.~Martel, P.~R.~Shapiro and S.~Weinberg,
Astrophys.\ J.\  {\bf 492}, 29 (1998)
[arXiv:astro-ph/9701099];
J.~Garriga, M.~Livio and A.~Vilenkin,
Phys.\ Rev.\  D {\bf 61}, 023503 (2000)
[arXiv:astro-ph/9906210].

\bibitem{Bousso:2006ev}
R.~Bousso,
Phys.\ Rev.\ Lett.\  {\bf 97}, 191302 (2006)
[arXiv:hep-th/0605263].

\bibitem{Bousso:2007kq}
R.~Bousso, R.~Harnik, G.~D.~Kribs and G.~Perez,
Phys.\ Rev.\  D {\bf 76}, 043513 (2007)
[arXiv:hep-th/0702115].

\bibitem{Hall:2007ja}
L.~J.~Hall and Y.~Nomura,
Phys.\ Rev.\  D {\bf 78}, 035001 (2008)
[arXiv:0712.2454 [hep-ph]].

\bibitem{DeSimone:2008bq}
A.~De Simone, A.~H.~Guth, M.~P.~Salem and A.~Vilenkin,
arXiv:0805.2173 [hep-th].

\bibitem{Bousso:2008hz}
R.~Bousso, B.~Freivogel and I.~S.~Yang,
arXiv:0808.3770 [hep-th].

\bibitem{Bousso:2000xa}
R.~Bousso and J.~Polchinski,
JHEP {\bf 0006}, 006 (2000)
[arXiv:hep-th/0004134];
S.~Kachru, R.~Kallosh, A.~Linde and S.~P.~Trivedi,
Phys.\ Rev.\  D {\bf 68}, 046005 (2003)
[arXiv:hep-th/0301240];
L.~Susskind,
arXiv:hep-th/0302219;
M.~R.~Douglas,
JHEP {\bf 0305}, 046 (2003)
[arXiv:hep-th/0303194].

\bibitem{Tegmark:2005dy}
M.~Tegmark, A.~Aguirre, M.~J.~Rees and F.~Wilczek,
Phys.\ Rev.\  D {\bf 73}, 023505 (2006)
[arXiv:astro-ph/0511774].

\bibitem{Rees:1977}
M.~J.~Rees and J.~P.~Ostriker,
Mon.\ Not.\ R.\ astr.\ Soc.  {\bf 179}, 541 (1977).

\bibitem{Silk:1977wz}
J.~Silk,
Astrophys.\ J.\  {\bf 211}, 638 (1977).

\bibitem{White:1978}
S.~D.~M.~White and M.~J.~Rees,
Mon.\ Not.\ R.\ astr.\ Soc.  {\bf 183}, 341 (1978).

\bibitem{Tegmark:1997in}
M.~Tegmark and M.~J.~Rees,
Astrophys.\ J.\  {\bf 499}, 526 (1998)
[arXiv:astro-ph/9709058].

\bibitem{Carr:1979sg}
B.~J.~Carr and M.~J.~Rees,
Nature {\bf 278}, 605 (1979).

\bibitem{Page:2006dt}
D.~N.~Page,
arXiv:hep-th/0610079.

\bibitem{Bousso:2006xc}
R.~Bousso and B.~Freivogel,
JHEP {\bf 0706}, 018 (2007)
[arXiv:hep-th/0610132].

\bibitem{Bousso:2007nd}
R.~Bousso, B.~Freivogel and I.~S.~Yang,
Phys.\ Rev.\  D {\bf 77}, 103514 (2008)
[arXiv:0712.3324 [hep-th]].

\end{thebibliography}
\end{document}